\documentclass[a4paper,11pt]{article}
\pdfoutput=1 
\usepackage{jheppub} 
\usepackage{amsfonts}
\usepackage{textcomp}
\usepackage{amssymb,amsmath}
\usepackage{hyperref}
\usepackage{comment}
\usepackage{graphicx}
\usepackage{multirow,booktabs,multirow}
\usepackage{braket}
\usepackage{bbold}
\usepackage{xcolor}
\usepackage[utf8]{inputenc}
\usepackage{cleveref}
\usepackage{eqparbox}
\usepackage{float}
\usepackage{cancel}
\usepackage{slashed}
\usepackage{physics}
\usepackage{bbm}
\usepackage{units}
\usepackage{bm}
\usepackage{tikz}
\usepackage[compat=1.1.0]{tikz-feynman}
\tikzfeynmanset{warn luatex=false}
\usepackage{tikz}
\usetikzlibrary{decorations.markings,decorations.pathmorphing}
\usetikzlibrary{calc}

\title{Dark Neutrons as Dark Matter: Collisions in Halos and Direct Detection from Dark CP Violation
}

\author[a]{Pablo Figueroa,}
\emailAdd{pablo.figueroa@ific.uv.es}

\author[a]{Camilo Garc\'ia-Cely,}
\emailAdd{camilo.garcia@ific.uv.es}

\author[b]{Giacomo Landini,}
\emailAdd{giacomo.landini@lnf.infn.it}

\affiliation[a]{Instituto de F\'isica Corpuscular (IFIC), Consejo Superior de Investigaciones
Cient\'ificas (CSIC) and Universitat de Val\`encia,  C/ Catedratico Jose Beltran 2, E-46980 Paterna, Spain}

\affiliation[b]{Istituto Nazionale di Fisica Nucleare, Laboratori Nazionali di Frascati, C.P. 13, 00044 Frascati, Italy}

\begin{document}

\abstract{
We consider confining gauge theories with a non-vanishing topological angle $\theta$, which induces CP-violating interactions among dark pions and dark baryons, with dark matter consisting of dark neutrons.  The $\theta$ term generates scalar pion--baryon couplings analogous to the CP-violating pion--nucleon interactions of QCD. These interactions give rise to an attractive long-range Yukawa potential mediated by the dark pions, whose strength is proportional to $\theta$. Since the dark pions are naturally light pseudo-Goldstone bosons, the resulting force can lead to sizable dark matter self-interactions, providing a simple and theoretically motivated realization of the self-interacting dark matter paradigm.

We also investigate the implications for direct detection in scenarios where the dark sector communicates with the Standard Model through a dark photon portal. The $\theta$ term induces dark electric dipole moments proportional to $\theta$, which couple directly to the electric fields of nuclei and can substantially enhance direct detection rates. We show how the underlying interactions shape the recoil spectra and discuss the possibility of identifying the CP-violating origin of the signal through its time dependence.  Finally, we analyze the interplay between dark matter self-interactions and direct detection signals, showing that both are controlled by the same CP-violating parameter. Our results demonstrate that the topological angle $\theta$ can play a central role in determining the phenomenology of composite dark matter. 
} 

\maketitle

\section{Introduction}

The identity of the dark matter (DM) remains one of the central open problems in cosmology and astroparticle physics.  A comprehensive overview of the current experimental status and theoretical developments can be found in Ref.~\cite{Cirelli:2024ssz}. Among the many proposed candidates, strongly coupled dark sectors provide a particularly compelling framework, largely inspired by the dynamics of Quantum Chromodynamics (QCD).  In particular, confining sectors with characteristic mass scales below the electroweak scale have attracted significant interest. In these scenarios, DM may arise as pseudo-Goldstone bosons associated with the spontaneous breaking of a dark flavor symmetry, similar to the pions in the Standard Model (SM), or as composite baryonic states analogous to the proton and neutron of the SM. For a recent review, see Ref.~\cite{Asadi:2026mip}, see also Refs.~\cite{Hochberg:2015vrg, Kuflik:2015isi,Bernal:2015bla, Bernal:2015xba, Bernal:2015ova,  Choi:2015bya, Choi:2016hid,Soni:2016gzf, Kamada:2016ois,  Bernal:2017mqb,Cline:2017tka, Choi:2017mkk,  Kuflik:2017iqs, Heikinheimo:2018esa, Choi:2018iit,Hochberg:2018rjs, Bernal:2019uqr, Choi:2019zeb,  Katz:2020ywn, Smirnov:2020zwf, Xing:2021pkb, Braat:2023fhn, Bernreuther:2023kcg,Garani:2021zrr,Fiorentino:2026dea}.

An important and largely unexplored aspect of these theories is the role played by CP violation induced by a non-vanishing topological angle $\theta$. In QCD, the $\theta$ term gives rise to  CP-violating phenomena such as electric dipole moments and modified meson--baryon interactions~\cite{Crewther:1979pi, Baluni:1978rf, Pich:1995bw}. Similar effects are generically expected in confining dark sectors. A particularly important observation~\cite{Crewther:1979pi,Shifman:1979if} is that a non-vanishing $\theta$ angle induces both cubic meson interactions and scalar meson--baryon couplings. While the former have been studied in the context of dark pion DM~\cite{Kamada:2017tsq, Garcia-Cely:2024ivo,Garcia-Cely:2025flv}, the implications of the CP-violating meson--baryon couplings for composite baryonic DM remain comparatively unexplored. This is the subject of this work. In analogy with the SM, such interactions read~\cite{Crewther:1979pi}
\begin{align}
\mathcal{L}_{\rm CPV} \sim \bar g_{\pi NN}\,\bar N N \pi \,,
\qquad
\bar g_{\pi NN} \propto \theta\,,
\end{align}
to be contrasted with the ordinary pseudoscalar couplings, $\mathcal{L} \sim g_{\pi NN}\,\bar N i\gamma_5 N \pi $. Here, $N$ represents a dark baryonic state, while $\pi$ is a dark pion pseudo-Goldstone boson. 
One of the main consequences of the CP-violating  coupling is the emergence of a long-range Yukawa interaction between the DM particles 
\begin{align}
V(r) \sim -\frac{\bar g_{\pi NN}^2}{4\pi}\frac{e^{-m_\pi r}}{r}\,.
\label{eq:yukawa}
\end{align}
By contrast, the ordinary pseudoscalar interaction proportional to $g_{\pi NN}$ does not generate such a potential. Instead, it gives rise to spin- and momentum-dependent interactions whose effects on DM self-scattering are negligible in astrophysical environments~\cite{Kahlhoefer:2017umn,Agrawal:2020lea}.
In contrast, as long as $m_\pi \ll m_{\rm DM}$, the potential in Eq.~\eqref{eq:yukawa} induces strong velocity-dependent scattering among DM particles, in agreement with the paradigm of self-interacting dark matter (SIDM)~\cite{Spergel:1999mh}. 

The SIDM paradigm has long been recognized as a mechanism for reducing the central densities of DM halos, thereby potentially alleviating tensions between the cold DM paradigm and observations on galactic and sub-galactic scales. Comprehensive reviews of SIDM, as well as alternative explanations of the small-scale structure problems, can be found in Refs.~\cite{Tulin:2017ara,Adhikari:2022sbh}. At the same time, consistency with observations of galaxy clusters and other large-scale systems requires the self-interaction cross section to decrease to $\sigma/m_{\rm DM}\lesssim 0.1\,{\rm cm}^2/{\rm g}$ at sufficiently large velocities. In conventional SIDM scenarios, obtaining velocity-dependent self-interaction cross sections typically requires the introduction of a light scalar inducing a Yukawa potential, with a small mass imposed as an additional assumption. In contrast, within the framework considered here, the light mediator arises naturally as a consequence of the pseudo-Goldstone nature of the dark pions. Moreover, the required velocity dependence emerges automatically from the Yukawa potential, whose strength is controlled by $\theta$. This provides a simple and theoretically motivated realization of velocity-dependent DM self-interactions.
The role of CP-violating interactions for fermionic DM self-interactions mediated by a light spin-0 particle was previously explored in Ref.~\cite{Kahlhoefer:2017umn}, where both the mediator mass and its CP-violating couplings are treated as free parameters. In the present framework, both ingredients originate from the dynamics of a confining dark sector.

A second important consequence of CP violation concerns direct detection experiments. We consider scenarios in which the dark sector communicates with the SM through a dark photon portal~\cite{Holdom:1985ag}. In the CP-conserving limit, direct detection is typically dominated by magnetic dipole interactions of the dark baryons. These interactions couple to nuclei through spin-orbit effects and are velocity suppressed. In contrast, a non-vanishing $\theta$ angle induces dark electric dipole moments proportional to $\theta$~\cite{Antipin:2015xia}
\begin{align}
d_E \propto \theta\,,
\end{align}
in close analogy with the neutron electric dipole moment in QCD~\cite{Baluni:1978rf,Crewther:1979pi}. 
We calculate in detail the direct detection rates and find that they can be substantially larger than in the CP-conserving case, making direct detection experiments sensitive probes of dark-sector CP violation. 

Taking these results together, we argue that the topological angle $\theta$ can play a central role in determining the phenomenology of composite DM.
With this in mind, this paper is organized as follows. In Sec.~\ref{sec:qcdlike} we introduce the effective low-energy description of QCD-like dark sectors with a non-vanishing topological angle and discuss their cosmological history along with the corresponding constraints. In Sec.~\ref{sec:sidm} we analyze the long-range Yukawa interaction induced by CP violation and discuss its implications for DM self-interactions and small-scale structure.  In Sec.~\ref{sec:dd} we investigate the direct detection phenomenology associated with dark electric dipole moments. In Sec.~\ref{sec:bench} we collect our results into a representative benchmark point.
Finally, we summarize our conclusions in Sec.~\ref{sec:conclusions}. In the Appendix, we provide additional discussions that complement the main text.

\section{DM as baryons with CP-violating interactions}\label{sec:sec2}

\label{sec:qcdlike}

\subsection{General setup}
We consider a dark sector with a gauge symmetry $SU(N_c)$ and $N_f$ light Dirac fermions $q$ transforming in the fundamental representation
\begin{equation}
	\mathcal{L} = -\frac{1}{4}F_{\mu\nu}F^{\mu\nu} + \frac{g^2\theta}{32\pi^2} F_{\mu\nu}\widetilde{F}^{\mu\nu} + \bar{q}i\slashed{D} q - \left(\bar{q}_L M q_R + \text{h.c.}\right)\,.
 \label{eq:L}
\end{equation}
We take the quark mass matrix to be
$M=\mathrm{diag}(m_1,\ldots,m_{N_f})$, with all masses real and nonzero. Then, 
the presence of the $\theta$ term breaks the otherwise conserved CP symmetry.   

Strong interactions confine at an energy scale $\Lambda$, and for small quark masses, the theory exhibits an approximate flavor symmetry that is spontaneously broken by the formation of a condensate. As in ordinary QCD, the resulting low-energy spectrum contains both baryons and pseudo-Goldstone bosons associated with the broken generators of the flavor symmetry. The lightest baryons have masses of order $\Lambda$, are stable  and constitute the DM of the theory. In contrast, the pseudo-Goldstone bosons are parametrically lighter than the confinement scale and therefore mediate the low-energy interactions among baryons. In the presence of a nonvanishing $\theta$ angle, these interactions acquire CP-violating contributions with distinctive phenomenological signatures, which constitute the main focus of this work.

\subsection{A model with \texorpdfstring{$SU(2)$}{SU(2)} flavor symmetry}\label{sec:su2}

Instead of discussing the general case, for which refer the reader to Appendix~\ref{sec:UVmodel}, we rely on the analogy with ordinary QCD, where the dynamics are well understood for $N_f=2$ and quark masses satisfying $m_1\sim m_2 \ll \Lambda$. Although this setup is intended merely as an illustration, it already contains all the essential ingredients associated with CP violation that are relevant for our discussion.

As in the SM, the relevant light-flavor sector is approximately described by an $SU(2)$ flavor symmetry acting on the two lighter quarks. 
The corresponding low-energy spectrum then contains a stable isospin doublet of nucleons and an isospin triplet of light pions,
\begin{equation}
\label{eq:doublettriplet}
N=
\begin{pmatrix}
p\\
n
\end{pmatrix},
\qquad
\bm{\pi}=
\begin{pmatrix}
\pi^1\\
\pi^2\\
\pi^3
\end{pmatrix}.
\end{equation}
While the pions are parametrically lighter than $\Lambda$, the nucleons have masses of order $\Lambda$. At low energies, whenever CP is not conserved, the interactions between nucleons and pions can be generically described by 
\begin{equation}\label{eq:baryonPionLag}
  \mathcal{L}^{(\pi NN)}
  =
  g_{\pi NN}\,\bar{N} i \gamma_{5} \bm{\sigma}\!\cdot\!\bm{\pi}\, N
  +
  \bar{g}_{\pi NN}\,\bar{N} \bm{\sigma}\!\cdot\!\bm{\pi}\, N \, ,
\end{equation}
where $\sigma^a$  are the Pauli matrices. In the limit $\theta\ll1$~\cite{Crewther:1979pi}
\begin{equation}\label{eq:gbar}
\bar{g}_{\pi NN}
=
-\frac{2 c\, m_1 m_2}{f_\pi(m_1+m_2)}\,\theta \, ,
\end{equation}
where $f_\pi$ denotes the pion coupling constant and $c$ is a dimensionless constant.\footnote{
In the SM, $c=
(M_{\Xi}-M_\Sigma)/(2m_s-m_u-m_d) = 0.7$.
} In the following, we assume that the pion--nucleon couplings $g_{\pi NN}$ and $\bar g_{\pi NN}$ are sufficiently small to justify the use of perturbation theory. In Appendix~\ref{app:chiralLag}, using Goldberger-Treiman relations,  we determine the implications of this on other parameters. Note that this assumption is not realized in the SM, where the CP-conserving coupling $g_{\pi NN}$ is much larger than unity and the corresponding phenomenology is dominated by strong dynamics. 

\subsection{The dark photon portal}\label{sec:darkphoton}

Motivated by direct-detection searches, and guided by the analogy with ordinary QCD, we also introduce a dark photon as a simple and well-motivated portal between the dark and visible sectors. Concretely, we extend the dark sector by introducing a dark $U(1)_{\rm{d}}$ gauge symmetry with gauge coupling $e_{\rm{d}}$. As the charges of the two lightest quarks, we adopt 
\begin{equation}
\label{eq:darkcharge}
Q_{\rm d}=\mathrm{diag}(2/3,-1/3),
\end{equation}
so that
the dark sector closely resembles ordinary QCD. The corresponding dark photon,
with mass $m_V$ and kinetic mixing $\epsilon$, is described by 
\begin{equation}
    \mathcal{L}_{\rm DP}=-\frac{1}{4}V_{\mu\nu}V^{\mu\nu}+\frac{1}{2}m_V^2V_\mu V^\mu+\frac{\epsilon}{2}V_{\mu\nu}B^{\mu\nu},
\end{equation}
where $B_{\mu\nu}$ is associated with the SM hypercharge. We leave the origin of the dark photon mass unspecified.

The nucleon doublet in Eq.~\eqref{eq:doublettriplet} decomposes into a  dark proton and dark neutron. The dark proton carries unit charge under $U(1)_{\rm{d}}$, while the dark neutron is neutral. Similarly, the pion triplet decomposes into a neutral state $\pi^0 = \pi^3$ and  two charged states $\pi^\pm$ $=(\pi^1\pm i\pi^2)/\sqrt{2}$.
 
The dark photon explicitly breaks the flavor symmetry through the dark charge assignments, generating positive radiative corrections to the masses of the charged states, rendering them slightly heavier than their neutral counterparts. Assuming $m_\pi \gtrsim m_V$ and  that the $U(1)_{\rm{d}}$ interaction is the dominant source of  mass splitting, we get 
\begin{align}\label{eq:split2}
\frac{m_p-m_n}{m_n}
    &\sim \frac{e_{\rm d}^2}{4\pi}
    \sim 0.08\,e_{\rm d}^2,
&
m_{\pi^\pm}^2-m_{\pi^0}^2
    &\simeq
    \frac{3e_{\rm d}^2 m_\rho^2}{8\pi^2}\ln 2\,,
\end{align}
The expression for the pion mass splitting is obtained by analogy with ordinary QCD~\cite{Contino:2010rs}, where $m_\rho \sim \Lambda$ denotes the characteristic mass scale of the vector resonances in the strongly interacting sector.\footnote{Corrections induced directly by the dark photon mass remain negligible as long as $m_V\ll \Lambda$ (see e.g. App.~B of Ref.~\cite{Balkin:2018tma}).}

\paragraph{Laboratory constraints.}

We will consider $m_{\pi^0} \gtrsim m_V$, so that the decay channels of the dark photon into dark sector states are kinematically closed, and its decay proceeds through kinetic mixing into visible SM final states. For the choice of parameters in Table~\ref{tab:benchmark-star}, which will be analyzed in detail throughout, the most constraining limits come from visible proton beam-dump searches in which the $V$ is produced in the target and decays into charged leptons, dominantly $V \rightarrow e^{+} e^{-}$. For the kinetic mixing $\epsilon$, we take values near the current upper bounds from BEBC \cite{Marocco:2020dqu, Barouki:2022bkt}, CHARM \cite{Boiarska:2021yho}, NA62 (dump mode) \cite{NA62:2023qyn, NA62:2023nhs} and NuCAL \cite{Blumlein:1990ay}, in order to maximize the direct-detection signal allowed by those constraints. 

\subsection{Cosmological history and constraints}

\label{sec:cosmo}

\begin{table}[t]
    \centering
    \renewcommand{\arraystretch}{1.3}
    \begin{tabular}{ccccccccc}
        \toprule
        $m_n$ & $m_{\pi^0}$ & $m_V$ & $g_{\pi NN}$ & $\bar g_{\pi NN}$ & $\epsilon$ & $e_{\rm d}$ & $\langle\sigma_V/m_n\rangle$ & $\langle\sigma_{\rm eff}/m_n\rangle$ \\
        (GeV) & (MeV) & (MeV) & & & & & (cm$^2$/g) & (cm$^2$/g) \\
        \midrule
        96.14 & 557.9 & 150.0 & 0.11 & 0.35 & $2\times10^{-5}$ & 1.0 & 3.65 & 1.8 \\
        \bottomrule
    \end{tabular}
    \caption{Benchmark point considered in this work. The viscosity cross section is averaged over a Maxwellian distribution with velocity dispersion $v_0=10~\mathrm{km\,s^{-1}}$, while the effective cross section, as introduced in Ref.~\cite{Yang:2022hkm}, corresponds to $\sigma_{\rm 1D}^{\rm eff}=5.14~\mathrm{km\,s^{-1}}$.} 
    \label{tab:benchmark-star}
\end{table}

The dark sector may thermalize with the SM plasma and track its temperature, or instead evolve independently, depending on the strength of the interactions mediated by the dark photon. For instance, for the choice of parameters corresponding to the benchmark point in Table~\ref{tab:benchmark-star}, we find that elastic scatterings of dark protons and charged dark pions with SM charged particles (leptons and SM pions) efficiently transfer energy between the two sectors, thereby establishing kinetic equilibrium at a common temperature. 

The observed DM cosmological relic abundance can be reproduced through different mechanisms. A particularly compelling scenario is the one in which the DM is asymmetric.
This is especially attractive in our framework, as the dark sector closely resembles the SM, where the abundance of baryons does arise from a primordial asymmetry. In the following, we focus on the asymmetric DM scenario, though remaining agnostic regarding the specific mechanism responsible for generating the primordial asymmetry. We briefly comment on a possible different cosmological history in App.~\ref{alternatives}.

Although the precise prediction  for the DM mass depends on the specific mechanism responsible for generating the asymmetry,
DM masses $m_n\sim \mathcal{O}(1-100)$ GeV are easily realized, especially in scenarios relating the dark and visible-sector asymmetries (see for instance~\cite{Petraki:2013wwa} and references therein). In this mass range, the annihilation processes $\bar{n}n\to\pi\pi$ (and analogously for $\bar{p}p$) are very efficient and erase the symmetric component of dark neutron DM (and dark protons)\footnote{The cross sections for $\bar{n}n(\bar{p}p)\to\pi\pi$ scale as $\sigma v\sim 1/\Lambda^2\sim 1/m_n^2$. As long as $m_n\ll 10$ TeV, the symmetric population of baryons is efficiently erased. Dark protons additionally annihilate through $U(1)_{\rm{d}}$ interactions as $\bar{p}p\to VV$.}.
Notice that the asymmetric component of dark protons (accompanied by an equal amount of $\pi^-$ to preserve charge neutrality) converts into dark neutrons via $p\pi^-\to n\pi^0$ processes, which efficiently reduce the dark proton-to-neutron abundance ratio $Y_p/Y_n$\footnote{If kinematically allowed, the decay $p\to n\pi^+$ also contributes to the depletion of protons.}. For the choice of parameters in Table~\ref{tab:benchmark-star}, we estimate the final value of the ratio taking into account the corresponding mass splitting, finding $Y_p/Y_n\lesssim 10^{-5}$, so that the DM is made of dark neutrons. Since the DM particle is neutral, scatterings off SM particles, via dark photon mediation, are loop-suppressed, thus possibly evading constraints from DM direct searches, as we will discuss in Sec.~\ref{sec:dd}. In addition, as DM has not annihilated since recombination, the corresponding indirect detection constraints are automatically evaded.
 
Next, let us  discuss the fate of the dark pions. The charged ones are stable, being the lightest particles charged under $U(1)_{\rm{d}}$. However, given the relatively large mass splitting, they efficiently convert to neutral pions through the process $\pi^+\pi^-\to \pi^0\pi^0$, which remains active until $T\simeq m_{\pi^\pm}/20$. Furthermore, they also convert to dark photons, $\pi^+\pi^-\to VV$, through $U(1)_{\rm{d}}$ interactions. Thus, the population of charged pions is  exponentially suppressed and its contribution to the DM abundance is negligible. For instance, for the choice of parameters corresponding to Table~\ref{tab:benchmark-star}, we estimate the abundance of dark pions to be $\lesssim 10^{-4}$ of the total DM abundance.
On the other hand, the neutral pions decay to dark photons via the triangle anomaly, which induces a $\pi^0V_{\mu\nu}\widetilde{V}^{\mu\nu}$ vertex, in complete analogy with the decay of the SM neutral pion to photons. We compute its lifetime, which is analogous to that of the SM neutral pion as
\begin{eqnarray}
    \tau_{\pi^0\to VV} & = & \left(\frac{3}{N_c}\right)^2\frac{1024\pi^5 f_\pi^2}{m_{\pi^0}^3e_{\rm{d}}^4}\left(1-\frac{4m_V^2}{m_{\pi^0}^2}\right)^{-1/2} \\ & \overset{m_V\ll m_{\pi^0}}  \simeq & \left(2\times  10^{-16}\text{ sec }\right)\left(\frac{1}{e_{\rm{d}}}\right)^{4}\left(\frac{558\text{ MeV}}{m_{\pi^0}}\right)\left(\frac{0.04}{m_{\pi^0}/f_\pi}\right)^{2}\,,
\end{eqnarray}
where  $f_\pi\sim \sqrt{N_c}\Lambda/4\pi\sim \sqrt{N_c}m_n/4\pi$, and in the last step we used $N_c=3$. Finally, the dark photon decays to SM particles. For instance, for the values in Table~\ref{tab:benchmark-star}, it decays to $e^+e^-$. We checked that its lifetime is much smaller than 1 sec, thus evading all the relevant constraints from Big Bang Nucleosynthesis (BBN).

\section{Implications for Halos and Small-Scale Structure}\label{sec:sidm}

Having discussed the cosmological history of the dark sector and the present-day composition of DM in halos, we now turn to DM scattering.
\subsection{Scattering of DM particles}

DM self-scattering in astrophysical halos provides one of the clearest probes of the CP-violating interactions among baryonic DM encoded in Eq.~\eqref{eq:baryonPionLag}. The tree-level non-relativistic cross section is given by 
\begin{equation}
\frac{d \sigma}{d \Omega}=\frac{ \bar g_{\pi NN}^{\,2} m_n^2 }{32\,\pi^2 \,m_{\pi^0}^4}\left[
\bar g_{\pi NN}^{\,2} \left( 1-\frac{m_n^2}{\,m_{\pi^0}^2} v^2\right)
+
\,\frac{1}{4}g_{\pi NN}^{\,2}
v^2
\right]+{\cal O}(v^4) 
\label{eq:tree}
\end{equation}
In the CP-conserving limit, $\bar g_{\pi NN}=0$, and pion exchange between DM baryons proceeds exclusively through the pseudoscalar interaction proportional to $g_{\pi NN}$. The resulting self-scattering cross section is suppressed by $v^4$, rendering it negligible in  DM halos, where $v\ll1$. Importantly, this conclusion is not altered by long-range dynamics: the pseudoscalar interaction remains perturbative and does not generate sizable nonperturbative enhancements, such as the resonant effects associated with Sommerfeld factors or near-threshold resonances~\cite{Agrawal:2020lea}.

One of the central conceptual points of this work is that the situation changes qualitatively in the presence of CP violation. For $\bar g_{\pi NN}\neq0$, the tree-level cross section in Eq.~\eqref{eq:tree} is no longer velocity suppressed. More importantly,  dark pion exchange in baryon scattering induces an attractive long-range force between DM particles. For sufficiently light mediators and low relative velocities, the scattering dynamics may enter a nonperturbative regime, even if $\bar g_{\pi NN}\ll 1$. As a result, collisions inside astrophysical halos may exhibit resonant and other long-range scattering effects.

As discussed in Sec.~\ref{sec:cosmo}, we assume that the dark neutron is the lightest baryonic state and that no antibaryonic component survives in present-day halos. Then, the long-range force between dark neutrons is governed by the Yukawa potential induced by $\pi^0$
\begin{equation}
\label{eq:yukawa_potential_simple}
V(r) =
-
\frac{\bar g_{\pi NN}^2}{4\pi}
\frac{e^{-m_{\pi^0} r}}{r}\,,
\end{equation}
as follows directly from Eq.~\eqref{eq:baryonPionLag}. See e.g. Ref.~\cite{Petraki:2016cnz} for details. 
While the CP-conserving interaction proportional to $g_{\pi NN}$ also induces a potential, it has a different radial and spin dependence, and its contribution to the scattering cross section is accurately captured within perturbation theory \cite{Agrawal:2020lea}, as stated above.  Since this is inconsequential in astrophysical halos as already mentioned, we neglect it in what follows.

To obtain the differential cross section, including the non-perturbative effects induced by the Yukawa potential in Eq.~\eqref{eq:yukawa_potential_simple}, one must solve the corresponding scattering problem by means of the Schrödinger equation. Since we consider dark  neutrons, i.e. asymmetric DM composed of identical spin-$1/2$ baryons with mass $m_n$, the cross section takes the form~\cite{1930RSPSA.126..259M,Landau:1991wop}
\begin{equation}
\frac{d\sigma}{d\Omega}
=
\frac{1}{4}
\left|f(\theta)+f(\pi-\theta)\right|^2
+
\frac{3}{4}
\left|f(\theta)-f(\pi-\theta)\right|^2\,,
\label{eq:identicalcross}
\end{equation}
where the coefficients $1/4$ and $3/4$ correspond to the statistical weights of the spin-singlet and spin-triplet channels, respectively. Here $f(\theta)$ is the scattering amplitude, which admits the partial-wave expansion
\begin{equation}
f(\theta)
=
\sum_{\ell=0}^{\infty}
(2\ell+1)
\left(
\frac{e^{2i\delta_\ell(k)}-1}{2ik}
\right)
P_\ell(\cos\theta)\,,
\label{eq:partialwaveexpansion}
\end{equation}
where $k=m_n v/2$ is the relative momentum and $\delta_\ell(k)$ denotes the phase shift of the $\ell$th partial wave. To determine the phase shifts, we follow Ref.~\cite{Chu:2019awd} and reformulate the Schrödinger equation as a first-order differential equation for the auxiliary function $\delta_{\ell,k}(r)$,
\begin{equation}
\label{camilo_eq}
\frac{d\delta_{\ell,k}(r)}{dr}
=
-k\,m_n\,r^2\,V(r)
\left[
\cos\!\big(\delta_{\ell,k}(r)\big)\,j_\ell(kr)
-
\sin\!\big(\delta_{\ell,k}(r)\big)\,n_\ell(kr)
\right]^2,
\end{equation}
where $j_\ell$ and $n_\ell$ are spherical Bessel functions, subject to the boundary condition  $\delta_{\ell,k}(0)=0$. The phase-shift is thus $\delta_\ell(k)=\delta_{\ell,k}(r\rightarrow\infty)$.

For studies of DM halos, it is convenient to consider the viscosity cross section~\cite{Tulin:2013teo}, $\sigma_V \equiv \int d\Omega\,\sin^2\theta\, d\sigma/d\Omega$,
which, upon using Eq.~\eqref{eq:identicalcross}, can be expressed in terms of the phase shifts as~\cite{Colquhoun:2020adl}
\begin{align}
\sigma_V
=
\frac{8\pi}{k^2}
\sum_{\ell=0}^{\infty}
&\left[
\frac{1}{4}\,\frac{(2\ell+1)(2\ell+2)}{4\ell+3}
\sin^2\!\left(\delta_{2\ell+2}\!-\!\delta_{2\ell}\right)
\right.
\nonumber\\
&\left.
+
\frac{3}{4}\,\frac{(2\ell+2)(2\ell+3)}{4\ell+5}
\sin^2\!\left(\delta_{2\ell+3}\!-\!\delta_{2\ell+1}\right)
\right].
\label{eq:sigmaVfermion}
\end{align}
Following Ref.~\cite{Tulin:2013teo},  we assume a Maxwellian distribution with velocity dispersion $v_0$ to model the DM velocity distribution in halos,

\begin{figure}
    \centering
  \includegraphics[width=0.8\linewidth]{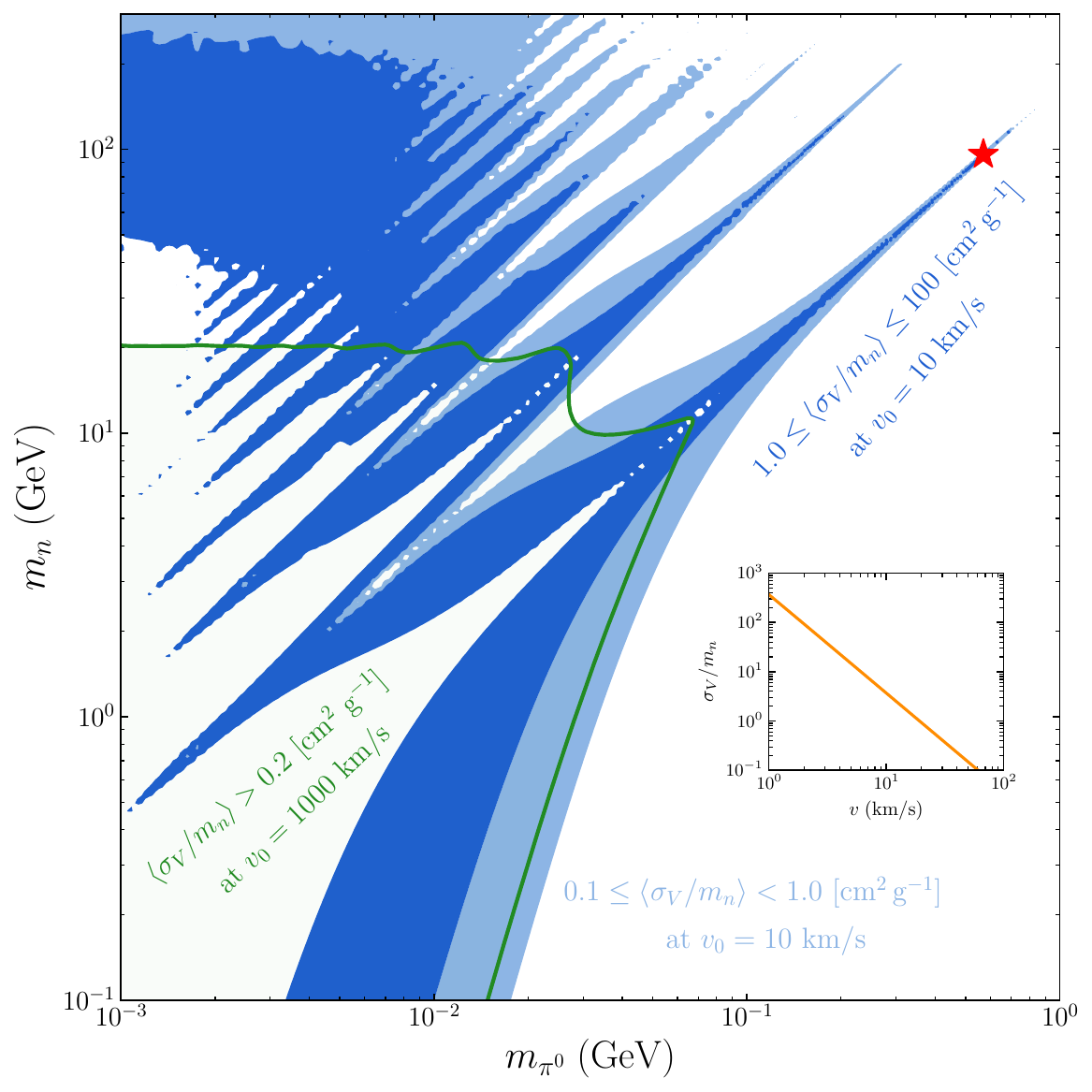}
    \caption{Velocity-averaged viscosity cross section per unit mass $\langle \sigma_V / m_n \rangle$ [cm$^2$ g$^{-1}$] for attractive interactions in the ($m_{\pi^0}$, $m_n$) plane, at fixed $\bar{g}_{\pi NN} = 0.35$. At small scales, the light blue band indicates cross-sections between $0.1 \leq \langle \sigma_V / m_n \rangle \leq 1.0$, while in dark blue $1.0 \leq \langle \sigma_V / m_n \rangle \leq 100 $. The cluster bound excludes parameter points in which $\langle \sigma_V / m_n \rangle  > 0.2 $ at cluster scales. The inset corresponds to the benchmark point indicated by the red star.}
    \label{fig:sidm_plot}
\end{figure}

\subsection{The SIDM paradigm}

The SIDM paradigm~\cite{Spergel:1999mh} has long attracted attention as a possible explanation for apparent discrepancies between observations and the predictions of collisionless CDM on small scales. 
Although differing in their details, these anomalies share a common feature: they suggest that DM halos are less centrally concentrated than predicted by collisionless simulations~\cite{Navarro:1995iw,Navarro:1996gj}. This behavior arises naturally in SIDM, where elastic scattering between DM particles redistributes energy through heat transport, transferring heat from the hotter outer regions toward the center and thereby reshaping the internal structure of halos, ultimately reducing the central densities. Comprehensive reviews and alternative explanations to the small-scale problems can be found in Refs.~\cite{Tulin:2017ara,Adhikari:2022sbh}. At the same time, the success of collisionless CDM in describing large-scale structure and galaxy clusters requires these interactions to become inefficient in high-velocity environments~\cite{Harvey:2015hha, Bondarenko:2017rfu, Harvey:2018uwf, Sagunski:2020spe, DES:2023bzs}. Taken together, these considerations point toward a velocity-dependent self-interaction cross section, with sizeable scattering rates in dwarf and galactic halos but substantially weaker interactions at cluster scales. More specifically, galaxy clusters, with characteristic velocities $v_0\sim1000~\mathrm{km\,s^{-1}}$, require $\sigma_V/m_{\rm DM}\lesssim0.2~\mathrm{cm^2\,g^{-1}}$, while dwarf galaxies and galaxies, where $v_0\sim10$--$100~\mathrm{km\,s^{-1}}$, favor $\sigma_V/m_{\rm DM}\gtrsim1~\mathrm{cm^2\,g^{-1}}$~\cite{Tulin:2017ara}.

As discussed in the previous section, the CP-violating pion exchange induced by a non-vanishing $\theta$ angle generates an attractive Yukawa interaction between dark baryons that is not subject to the suppressions affecting the CP-conserving case. A natural question is then whether the resulting self-scattering cross sections can reproduce the emerging astrophysical picture described above. To address this question, Fig.~\ref{fig:sidm_plot} displays the velocity-averaged viscosity cross section per unit dark neutron mass in the $(m_{\pi^0},m_n)$ plane for a representative value of the coupling, $\bar g_{\pi NN}=0.35$. The highlighted regions indicate the range of cross sections typically associated with observable effects in small-scale halos, together with the constraints arising from cluster systems. While a systematic comparison with observations across the different astrophysical scales lies beyond the scope of the present work, Fig.~\ref{fig:sidm_plot} already demonstrates that the range of self-scattering cross sections suggested by current observations can be naturally achieved within our framework. The star corresponds to the point in Table~\ref{tab:benchmark-star}.

So far we have focused on the region of parameter space yielding self-interaction cross sections of order $\sigma_V/m_{\rm DM}\sim1~\mathrm{cm^2\,g^{-1}}$. However, Fig.~\ref{fig:sidm_plot} and its inset show that our model naturally accommodates considerably larger self-interaction cross sections, possibly reaching $\sigma_V/m_{\rm DM}\sim100~\mathrm{cm^2\,g^{-1}}$ at low velocities. Such large cross sections can drive SIDM halos into the gravothermal-collapse regime over cosmological timescales. This regime has recently attracted renewed attention because it may account for the existence of unusually compact and dense halos reported in a growing number of astrophysical studies; see, for example, Refs.~\cite{Yu:2025tmp,Vegetti:2026mmx,Li:2025kpb}. Gravothermal collapse results from the continued inward transport of heat, which eventually reverses the initial decrease of the central density and causes the inner region of the halo to contract, leading to a rapid increase in its density. Whether a halo reaches this stage depends not only on the self-interaction cross section but also on its mass, concentration, assembly history, and environment. A detailed comparison with observations lies beyond the scope of the present work. Nevertheless, Fig.~\ref{fig:sidm_plot} shows that the parameter space of our model naturally extends into the regime of large self-interaction cross sections relevant for gravothermal collapse.

Up to now, our discussion of the role of the topological angle has relied on perturbation theory, assuming $\bar g_{\pi NN},\,g_{\pi NN}\ll1$. While definitive conclusions beyond this regime are difficult to draw, low-energy scattering can be systematically studied within the framework of effective range theory~\cite{Chu:2019awd, Bethe:1949yr}. Using this approach together with lattice QCD results for low-energy nucleon--nucleon scattering, Ref.~\cite{Cline:2022leq} showed that, in the CP-conserving limit, obtaining a sizable velocity dependence in the scattering cross section is rather non-generic, occurring only within a narrow region of parameter space. Indeed, in the SM neutron--proton scattering displays a velocity-dependent cross section, although this behavior is well known not to originate from the pion-mediated Yukawa interaction but rather from the existence of the deuteron~\cite{Chu:2019awd,Tulin:2017ara}, whose unusually small binding energy enhances the low-energy scattering amplitude, yielding self-interaction cross sections of order a few $\mathrm{cm^2\,g^{-1}}$ at low velocities and a suppression around $10^3~\mathrm{km\,s^{-1}}$. In this context, our perturbative analysis suggests that a non-vanishing $\theta$ angle opens up additional regions of parameter space in which the desired velocity dependence can be realized with the appropriate order of magnitude beyond the perturbative regime.

\section{Direct Detection induced by  CP-violating interactions}\label{sec:dd}

\begin{figure}[t!]
\centering
\begin{tikzpicture}

  \begin{scope}[xshift=-4.5cm]

    \node (L_chi_in)  at (-2.5,  1.25) {\(n\)};
    \node (L_chi_out) at ( 2.5,  1.25) {\(n\)};

    \node (L_SM1) at (-2.5,-3.0) {\(\mathcal{N}\)};
    \node (L_SM2) at ( 2.5,-3.0) {\(\mathcal{N}\)};

    \coordinate (L_vLT1)  at (-1, 0);
    \coordinate (L_vLT2)  at ( 1, 0);
    \coordinate (L_vLTD)  at ( 0,-1);
    \coordinate (L_vLPD)  at ( 0,-3);
    \coordinate (L_X)     at ( 0,-2);
    \coordinate (L_SMMID) at ( 0,-3);

    \draw[/tikzfeynman/fermion] (L_chi_in) -- (L_vLT1);
    \draw[/tikzfeynman/fermion] (L_vLT2) -- (L_chi_out);

    \draw[/tikzfeynman/scalar] (L_vLT1) -- (L_vLT2)
      node[midway, above] {\(\pi^{\pm}\)};

    \draw[/tikzfeynman/fermion] (L_vLT1) -- (L_vLTD)
      node[midway, below left] {\(p\)};

    \draw[/tikzfeynman/anti fermion] (L_vLT2) -- (L_vLTD)
      node[midway, below right] {\(p\)};

    \draw[/tikzfeynman/photon] (L_vLTD) -- (L_vLPD)
      node[midway, right] {\(\epsilon\)};

    \draw[/tikzfeynman/fermion] (L_SM1) -- (L_SMMID);
    \draw[/tikzfeynman/fermion] (L_SMMID) -- (L_SM2);

    \node[circle, draw=black, fill=black, inner sep=0pt, minimum size=2pt]
      at (L_vLT1) {};
    \node[circle, draw=black, fill=black, inner sep=0pt, minimum size=2pt]
      at (L_vLT2) {};

    \node at (L_X) {\(\times\)};

    \node at (-0.25,-1.575) {\(V\)};
    \node at (-0.25,-2.50) {\(\gamma\)};

  \end{scope}

  \begin{scope}[xshift=3.0cm]

    \node (R_chi_in)  at (-2.5,  1.25) {\(n\)};
    \node (R_chi_out) at ( 2.5,  1.25) {\(n\)};

    \node (R_SM1) at (-2.5,-3.0) {\(\mathcal{N}\)};
    \node (R_SM2) at ( 2.5,-3.0) {\(\mathcal{N}\)};

    \coordinate (R_vLT1)  at (-1, 0);
    \coordinate (R_vLT2)  at ( 1, 0);
    \coordinate (R_vLTD)  at ( 0,-1);
    \coordinate (R_vLPD)  at ( 0,-3);
    \coordinate (R_X)     at ( 0,-2);
    \coordinate (R_SMMID) at ( 0,-3);

    \draw[/tikzfeynman/fermion] (R_chi_in) -- (R_vLT1);
    \draw[/tikzfeynman/fermion] (R_vLT2) -- (R_chi_out);

    \draw[/tikzfeynman/fermion] (R_vLT1) -- (R_vLT2)
      node[midway, above] {\(p\)};

    \draw[/tikzfeynman/scalar] (R_vLT1) -- (R_vLTD)
    node[midway, below left, xshift=+0.75mm, yshift=+0.75mm] {\(\pi^{\pm}\)};

    \draw[/tikzfeynman/scalar] (R_vLT2) -   - (R_vLTD)
    node[midway, below right, xshift=-0.75mm, yshift=+0.75mm] {\(\pi^{\pm}\)};

    \draw[/tikzfeynman/photon] (R_vLTD) -- (R_vLPD)
      node[midway, right] {\(\epsilon\)};

    \draw[/tikzfeynman/fermion] (R_SM1) -- (R_SMMID);
    \draw[/tikzfeynman/fermion] (R_SMMID) -- (R_SM2);

    \node[circle, draw=black, fill=black, inner sep=0pt, minimum size=2pt]
      at (R_vLT1) {};
    \node[circle, draw=black, fill=black, inner sep=0pt, minimum size=2pt]
      at (R_vLT2) {};

    \node at (R_X) {\(\times\)};

    \node at (-0.25,-1.575) {\(V\)};
    \node at (-0.25,-2.50) {\(\gamma\)};

  \end{scope}

\end{tikzpicture}

\caption{Loop-induced DM-nucleons scattering mediated by the dark photon.}
\label{fig:dark-neutron-em-vertex}
\end{figure}
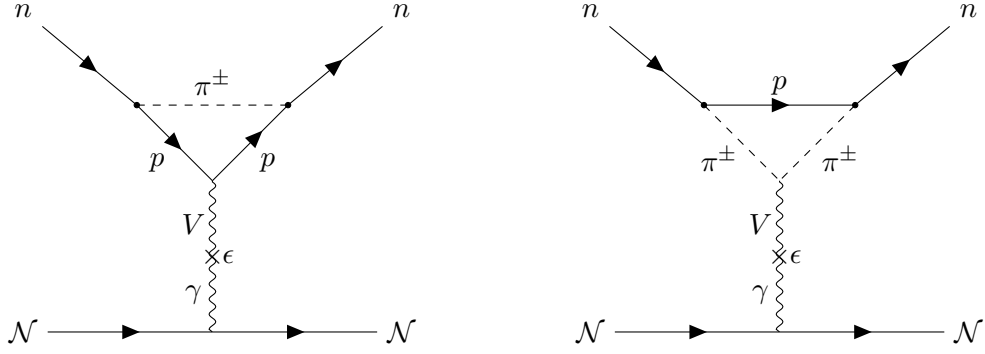

Having established that DM consists of dark neutrons interacting with the SM fields through a kinetically mixed dark photon, we are now in a position to investigate the resulting direct-detection phenomenology. DM electromagnetic properties have been extensively studied in the context of direct detection~\cite{Sigurdson:2004zp,Okawa:2019arp,Barger:2010gv,Antipin:2015xia}, since they can induce nuclear or electron recoils for neutral or millicharged DM~\cite{Banks:2010eh,DelNobile:2014eta,Hambye:2021xvd,Kuwahara:2026hpr}. We begin by considering scattering off an individual SM nucleon. The resulting DM--nucleon interaction will later be promoted to the nuclei through the standard non-relativistic effective field theory and the corresponding nuclear response functions.

In our framework, the dark neutron couples to the current, $J^\mu$, associated with the dark charge in Eq.~\eqref{eq:darkcharge}, while the SM nucleon couples through the ordinary electromagnetic current, $J^\mu_{\rm EM}$. Kinetic mixing mediates the interaction between these currents. The corresponding elastic scattering diagrams are shown in Fig.~\ref{fig:dark-neutron-em-vertex}, where the dark neutron is denoted by $n$ and the SM nucleons by $\mathcal{N}$. The corresponding scattering amplitude is 
\begin{align}
\mathcal{M}
&=
-i\epsilon\,
\langle n|J^\mu|n\rangle
\left(
\frac{i g_{\mu\nu}}
{q^2+m_{V}^2}
\right)
\langle \mathcal N|J^\nu_{\rm EM}|\mathcal N\rangle,
\label{eq:amplitude}
\end{align}
Since the visible current is well known~\cite{Weinberg:1995mt}, we focus on parametrizing the dark electromagnetic current. 

\subsection{Dark electromagnetic form factors}

Neutrality only fixes the total charge of the dark neutron to zero,  while its composite nature gives rise to non-trivial electromagnetic form factors, as in the case of the ordinary neutron.
In a CP violating theory, they can be parameterized as
\begin{equation}
\bra{n(p^{\prime})} J^\mu \ket{n(p)}
=
\bar{u}(p^{\prime})\,\Gamma^{\mu}(q)\,u(p),
\label{eq:gamma}
\end{equation}

where $q=p^{\prime}-p$ is the momentum transferred by the dark photon and~\cite{Kubis:2000zd, Scherer:2005ri} 
\begin{equation}
    \Gamma^{\mu}(q) = \gamma^{\mu} F_{E}(q^2) + \frac{i \sigma^{\mu \nu} q_{\nu}}{2 m_{n}}  F_{M} (q^2) + \frac{\sigma^{\mu \nu} q_{\nu} \gamma^{5}}{2 m_{n}}  F_{D}(q^2) + \frac{(\gamma^\mu q^2 - 2m_{n} q^\mu) \gamma^5}{m_{n}^2}  F_{A}(q^2)
    \label{eq:formfactors}
\end{equation}
where $F_E$, $F_M$, $F_D$, and $F_A$ are the charge, magnetic-dipole, electric-dipole, and anapole form factors, respectively. Since direct-detection experiments probe small momentum transfers, we are interested only in their leading-order expansion around $q^2=0$. Let us discuss them separately 
\begin{itemize}
\item
As the dark neutron carries no charge,   the leading low-momentum contribution is $F_E(q^2) = q^2 b_n/6$ where $b_n$ is the conventional charge radius squared. This quantity cannot be calculated perturbatively, but it can be estimated on dimensional grounds as $b_{n} \sim e_{\textrm{d}}/ \Lambda^2 \sim e_{\textrm{d}} /m_{n}^2$ \cite{Pich:1995bw}.
\item Next, $ F_M(0)= 2m_n \mu_n $, where $\mu_n$ is the magnetic dipole moment (MDM). As for the charge radius, it is non-calculable and we estimate it in units of the dark Bohr magneton $ \mu_{n} \sim e_{\textrm{d}}/(2 m_{n})$ \cite{Jenkins:1992pi, Durand:1997ya}.

\item Similarly, $ F_D(0) =2m_n d_n$, where $d_n$ is the electric dipole moment (EDM). Unlike the charge radius and magnetic dipole, in a CP-conserving theory, it must vanish because the electromagnetic current is a polar vector. Then, the CP-violating pion-nucleon coupling generates a calculable contribution to $d_n$ 
from the loop diagrams of Fig. \ref{fig:dark-neutron-em-vertex},
\begin{equation}\label{dchi_muchi}
    d_{n} \simeq \frac{e_{\textrm{d}}}{2 m_{n}}\frac{g_{\pi NN} \bar{g}_{\pi NN}}{2 \pi^2} \log \left( \frac{m_{n}}{m_{\pi^{\pm}}}\right)\,,
\end{equation}
where we take the limit $m_\pi\ll m_n$. This agrees with the analogous result for the SM neutron \cite{Crewther:1979pi, Pich:1991fq}.
\item The  anapole form factor $F_A$ is also CP-violating. Nonetheless, the contribution from the pion loops in Fig.~\ref{fig:dark-neutron-em-vertex} vanishes, giving a subdominant effect~\cite{Maekawa:2000qz}. 
\end{itemize}

\begin{table}[t!]
\centering
\renewcommand{\arraystretch}{1.8}
\begin{tabular}{|c|c|c|c|}
\hline
\multicolumn{3}{|c|}{\textbf{Operator}} 
& 
\textbf{Wilson Coefficient}
\\
\hline
\hline

$O_1^{\mathcal{N}}$
&
$1_n 1_{\mathcal{N}}$
&
CP-even
&
$\frac{e c_W \epsilon Q_{\mathcal{N}}}{2 m_{n}} \frac{q^2}{q^2 + m_V^2} \mu_{n} + e c_W \epsilon Q_{\mathcal{N}} \frac{q^2}{q^2 + m_V^2} b_{n}$
\\

$O_4^{\mathcal{N}}$
&
$\bm{S}_n \cdot \bm{S}_{\mathcal{N}}$
&
CP-even
&
$\frac{e c_W \epsilon g_{\mathcal{N}}}{m_{\mathcal{N}}} \frac{q^2}{q^2 + m_{V}^2} \mu_{n}$
\\

$O_5^{\mathcal{N}}$
&
$i \bm{S}_n \cdot
\left(
\dfrac{\bm{q}}{m_{\mathcal{N}}} \times \bm{v}^{\,\perp}
\right)$
&
CP-even
&
$\frac{2 e c_W \epsilon Q_{\mathcal{N}} m _{\mathcal{N}}}{q^2 + m_{V}^{2}} \mu_{n}$
\\

$O_6^{\mathcal{N}}$
&
$\left(
\bm{S}_n \cdot \dfrac{\bm{q}}{m_{\mathcal{N}}}
\right)
\left(
\bm{S}_{\mathcal{N}} \cdot \dfrac{\bm{q}}{m_{\mathcal{N}}}
\right)$
&
CP-even
&
$- \frac{e c_W \epsilon g_{\mathcal{N}} m _{\mathcal{N}}}{ q^2 + m_{V}^{2}} \mu_{n}$
\\

$O_{11}^{\mathcal{N}}$
&
$i \bm{S}_n \cdot \dfrac{\bm{q}}{m_\mathcal{N}}$
&
CP-odd
&
$\frac{2 e c_W \epsilon Q_{\mathcal{N}} m_{\mathcal{N}}}{q^2 + m_{V}^{2}} d_{n}$
\\ 
\hline
\end{tabular}
\caption{Relevant NREFT operators. Here $c_W$ denotes the cosine of the weak angle,  $Q_{\mathcal{N}}$ the electric charge of the SM proton and neutron, while $g_{\mathcal{N}}$ their corresponding $g-$factors. 
}
\label{tab:dipole_nonrelativistic_operators}
\end{table}

\paragraph{Non-relativistic expansion.} Since we are interested in elastic scattering of DM off nucleons, the relevant process involves one incoming and one outgoing dark neutron together with one incoming and one outgoing nucleon. At the low momentum transfers characteristic of direct-detection experiments, these interactions can be systematically described within the non-relativistic effective field theory (NREFT) framework. Following the standard operator classification of Refs.~\cite{Fitzpatrick:2012ix,Anand:2013yka}, the DM-nucleon scattering amplitude in Eq.~\eqref{eq:amplitude} can be written as\footnote{The coefficients $c_i^{\mathcal{N}}$ do not include relativistic external-state normalization factors.
\label{foot}
}
\begin{equation}\label{nr_lagrangian}
    \mathcal{M}
    = 
    \sum_{i}\sum_{\mathcal N}
    c_i^{\mathcal N}(q^2,\bm{v}_\perp^2)\,
    \bra{\mathcal{N}}\mathcal{O}_i^{\mathcal N} \ket{\mathcal{N}},
\end{equation}
where $\mathcal{O}_i^{\mathcal N}$ denotes the standard basis of non-relativistic operators, and the Wilson coefficients $c_i^{\mathcal N}$ encode the underlying particle physics. The operators are constructed from the independent kinematic quantities $(\bm q,\bm v_\perp,\bm S_n,\bm S_{\mathcal N})$, where $\bm S_n$ ($\bm S_{\mathcal N}$) denotes the dark-neutron (nucleon) spin operator and
$\bm v_\perp=\bm v+\bm q/(2\mu_T)$
is the transverse relative velocity, with $\mu_T$ being the DM-nucleus reduced mass~\cite{Fitzpatrick:2012ix,Anand:2013yka}.
The non-zero Wilson-coefficients, summarized in Table~\ref{tab:dipole_nonrelativistic_operators}, are obtained by expanding Eq.~\eqref{eq:amplitude} in the non-relativistic limit and matching the resulting amplitude onto the operator basis $\mathcal{O}_i^{\mathcal N}$.  Our results agree with those in Ref.~\cite{Kuwahara:2026hpr}. See also~\cite{Hambye:2021xvd, Ibarra:2024mpq}.

\subsection{Recoil rates}

\begin{figure}[t!]
    \centering
    \includegraphics[width=0.8\linewidth]{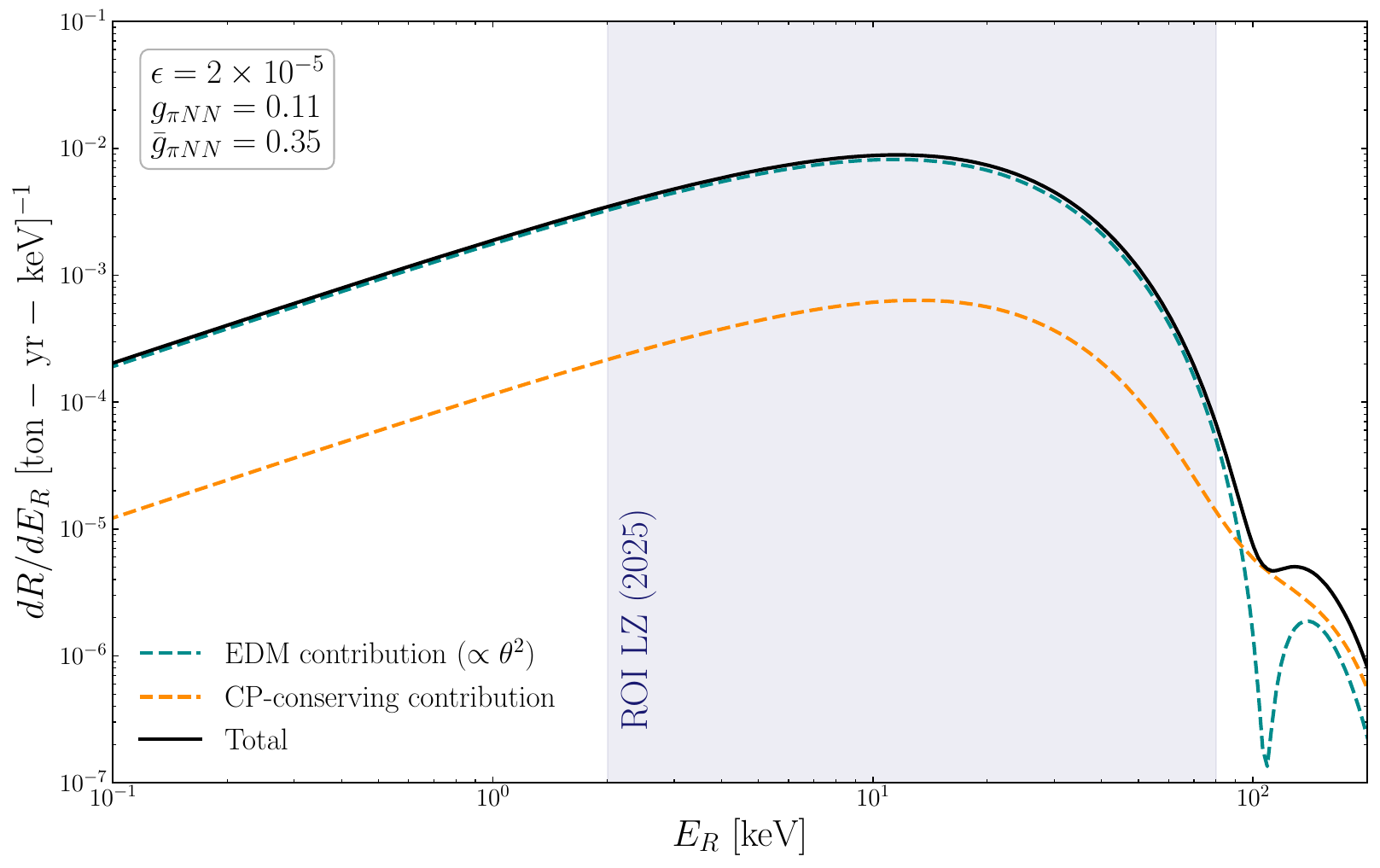}
    \caption{Differential event rate as a function of the nuclear recoil energy $E_R$ for the benchmark point considered in the text.} 
    \label{fig:rates}
\end{figure}

The DM-nucleus cross-section differential in recoil energy $E_R$ is~\cite{Fitzpatrick:2012ix}

\begin{equation}\label{diff_x_sec}
    \frac{d \sigma^T}{d E_R} = \frac{m_T}{2 \pi v^2} \left[ \frac{1}{(2 j_n + 1)} \frac{1}{(2j_{T}+1)} \sum_{\textrm{spins}} |\mathcal{M}^{T}|^2\right]\,
\end{equation}
with the same normalization adopted in Table~\ref{tab:dipole_nonrelativistic_operators} and Eq.~\eqref{nr_lagrangian},  consistent with footnote \ref{foot}.
Here $m_T$ ($j_T$) denotes the target nucleus mass (spin). The scattering amplitude off the target, $|\mathcal{M}^T|^2$, is obtained from the non-relativistic expansion described above as in Ref.~\cite{Fitzpatrick:2012ix}.
Once the cross section is obtained, the differential event rate of DM scattering off a nucleus with nuclear recoil $E_R$ is given by
\begin{equation}\label{rate}
\frac{dR}{dE_R}
= N_{T} \frac{\rho_n}{m_n} \int_{v > v_{\textrm{min}}} \frac{d\sigma^T}{dE_R} v  f_{\oplus}(\bm{v},\bm{v}_{E}(t)) \ d^3\bm{v}
\end{equation}
where $\rho_n \simeq 0.3\,\mathrm{GeV/cm^3}$ is the local DM density, $N_T$ the total number of nuclear targets $T$ in the detector and $f_{\oplus}(\bm{v},\bm{v}_{E}(t))$ 
the DM velocity distribution in the Earth frame. The minimum velocity required to produce a recoil with energy $E_{R}$ is given by $v_{\textrm{min}} = \sqrt{m_T E_R / (2 \mu_T^2)}$. For the numerical evaluation of Eq.~\eqref{rate}, we implement the Wilson coefficients and operators from Table~\ref{tab:dipole_nonrelativistic_operators} in \textsc{WimPyDD} \cite{Jeong:2021bpl}, which performs the spin average in Eq.~\eqref{diff_x_sec} accounting for the DM and nuclear response functions, as introduced in Ref.~\cite{Fitzpatrick:2012ix}. Unless stated otherwise, we adopt its default choices, including astrophysical parameters. In particular, for a Xenon target, the current implementation accounts for the natural abundances of the Xe isotopes, including their nuclear spins, and isotope-dependent response functions. 

In Fig.~\ref{fig:rates} we show the differential recoil spectra for the benchmark point specified in Table~\ref{tab:benchmark-star}. The total rate (solid black) closely follows the EDM contribution (dashed cyan), illustrating that the CP-violating component dominates once the CP-odd pion nucleon coupling $\bar{g}_{\pi NN}$ is taken somewhat larger than $g_{\pi NN}$.   For reference, the blue-shaded band indicates the  LZ (2025) region of interest (ROI) \cite{LZ:2024zvo}. 

\paragraph{Individual Operator Contributions and Interference.} 

The origin and relative size of each operator in Table \ref{tab:dipole_nonrelativistic_operators} can be qualitatively understood as follows.  In the target rest frame, the electric and magnetic fields of the nucleus are given by $\bm{E}=-\nabla\Phi$ and $\bm{B}=0$, where $\Phi$ is the scalar potential generated by the nuclear charge. In momentum space, $\nabla\rightarrow i\bm{q}$, so the interaction between the DM electric dipole moment ($d_n$) and the electric field of the target nucleus is proportional to $\bm{S}_n\cdot\bm{q}$, corresponding to the operator $\mathcal{O}_{11}$. On the other hand, the velocity-dependent operator $\mathcal{O}_5$ arises from the coupling of the magnetic dipole moment ($\mu_n$) to the magnetic field. Although the nuclear charge generates only an electric field in the laboratory frame, a magnetic field is induced in the DM rest frame by the Lorentz boost, $\bm{B}\sim\bm{E}\times\bm{v}$. This gives rise to the $\mathcal{O}_5$ structure, $\bm{S}_n\cdot(\bm{q}\times\bm{v}^{\perp})$. Additionally, the DM magnetic dipole moment ($\mu_n$) interacts directly with the nuclear magnetic dipole moment ($\mu_{\mathcal N}$). The resulting magnetic dipole--dipole interaction is proportional to $q^2(\bm{S}_n\cdot\bm{S}_{\mathcal N})-(\bm{S}_n\cdot\bm{q})(\bm{S}_{\mathcal N}\cdot\bm{q})$, which maps onto the operators $\mathcal{O}_4$ and $\mathcal{O}_6$. Thus, the magnetic dipole contribution separates into a dipole--charge interaction, associated here with $\mathcal{O}_{5}$, and a dipole--dipole interaction, associated with $\mathcal{O}_{4}$ and $\mathcal{O}_{6}$. See \cite{Barger:2010gv, DelNobile:2017fzy}. The spin-independent operator $\mathcal{O}_{1}$ receives subleading contributions from both the magnetic dipole and the charge radius, which are $q^2$-suppressed~\cite{Hambye:2021xvd}.

\begin{figure}[t!]
    \centering
    \includegraphics[width=0.81\linewidth]{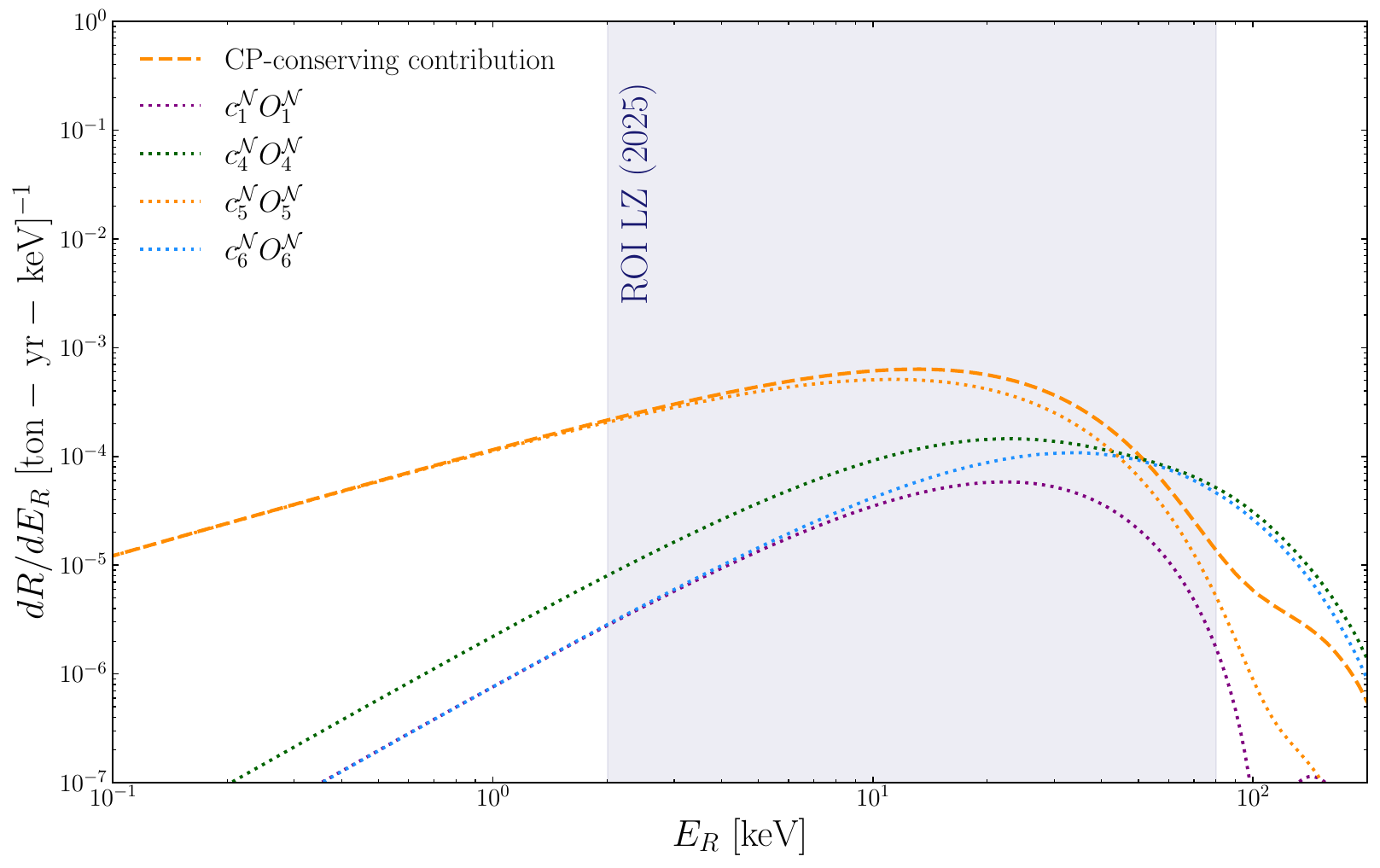}
    \includegraphics[width=0.81\linewidth]{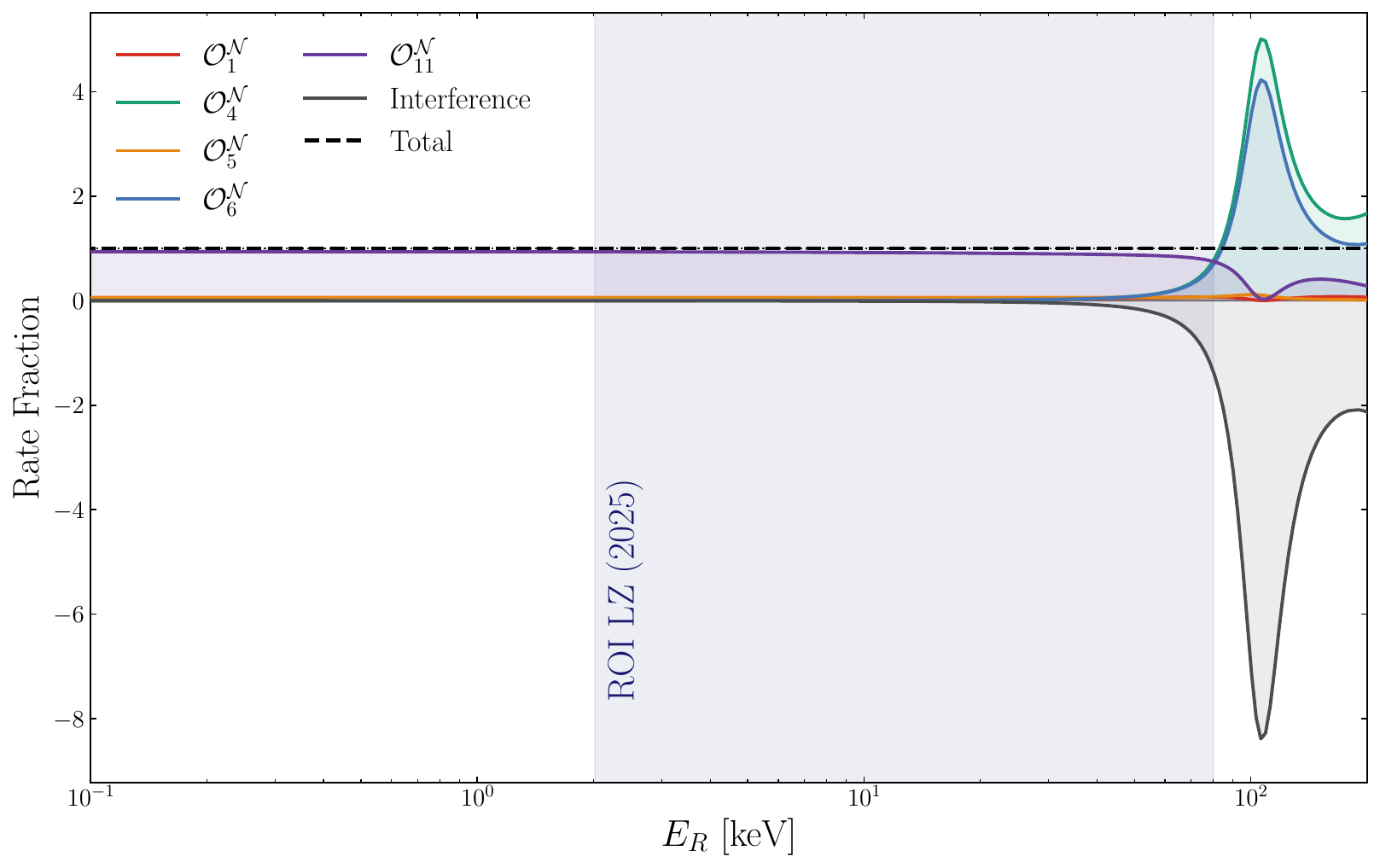}
    \caption{\textit{Top panel:} Individual $c_{i}^{\mathcal{N}}O_{i}$ contribution to the CP-conserving differential rate. \textit{Bottom panel:} Rate fraction of the individual operators, and the interference terms. We take the parameters of the benchmark point in Table~\ref{tab:benchmark-star}.}
    \label{fig:operator_contribution}
\end{figure}

The upper panel of Fig.~\ref{fig:operator_contribution} shows the differential recoil spectrum of the CP-conserving contribution (dashed orange) together with the individual combination $c_i^{\mathcal{N}}\mathcal{O}_{i}$ (dotted color) contributing to it. The hierarchy among the different contributions follows from both the Wilson coefficient and the nuclear responses. Over most of the LZ (2025) analysis window the rate is dominated by $c_5^{\mathcal{N}}\mathcal{O}_{5}$. The operators $\mathcal{O}_1$ and $\mathcal{O}_5$ both arise from the coupling of the dark-neutron magnetic dipole to the electric charge of the target nucleon, with the $\mathcal{O}_1$ contributions being $q^2$-suppressed. By contrast, $\mathcal{O}_4$ and $\mathcal{O}_6$ probe instead nuclear spin responses, which are not coherently enhanced in Xenon. As a result, $c_5^{\mathcal{N}}\mathcal{O}_{5}$ gives the leading contribution, despite its velocity dependence. The full recoil spectrum might be decomposed as the sum of the individual contributions and their corresponding interference terms, 
\begin{equation}\label{decomposition}
    \frac{dR}{dE_R} = \sum_i \frac{dR_i}{dE_R} + \sum_{i < j}\frac{dR^{\rm{Int}}_{ij}}{dE_R}\,.
\end{equation}
The rate fractions shown in the bottom panel of Fig.~\ref{fig:operator_contribution} are therefore the corresponding contributions in Eq.~\eqref{decomposition} normalized to the full differential rate. For the benchmark point in Table~\ref{tab:benchmark-star}, the total prediction is largely dominated by the CP-odd operator $\mathcal{O}_{11}$, while the leading subdominant contribution is $\mathcal{O}_{5}$. At larger recoil energies the fractions associated with $\mathcal{O}_4$ and $\mathcal{O}_5$, and their interference become visible, as their $q^2$ suppression is gradually lifted for increasing $E_R$. Both panels indicate that for the phenomenology of the LZ (2025) ROI, the CP-conserving contribution is well captured by the velocity-dependent $\mathcal{O}_5$.

\subsection{Direct detection limits}

\begin{figure}[t!]
    \centering    
    \includegraphics[width=0.81\linewidth]{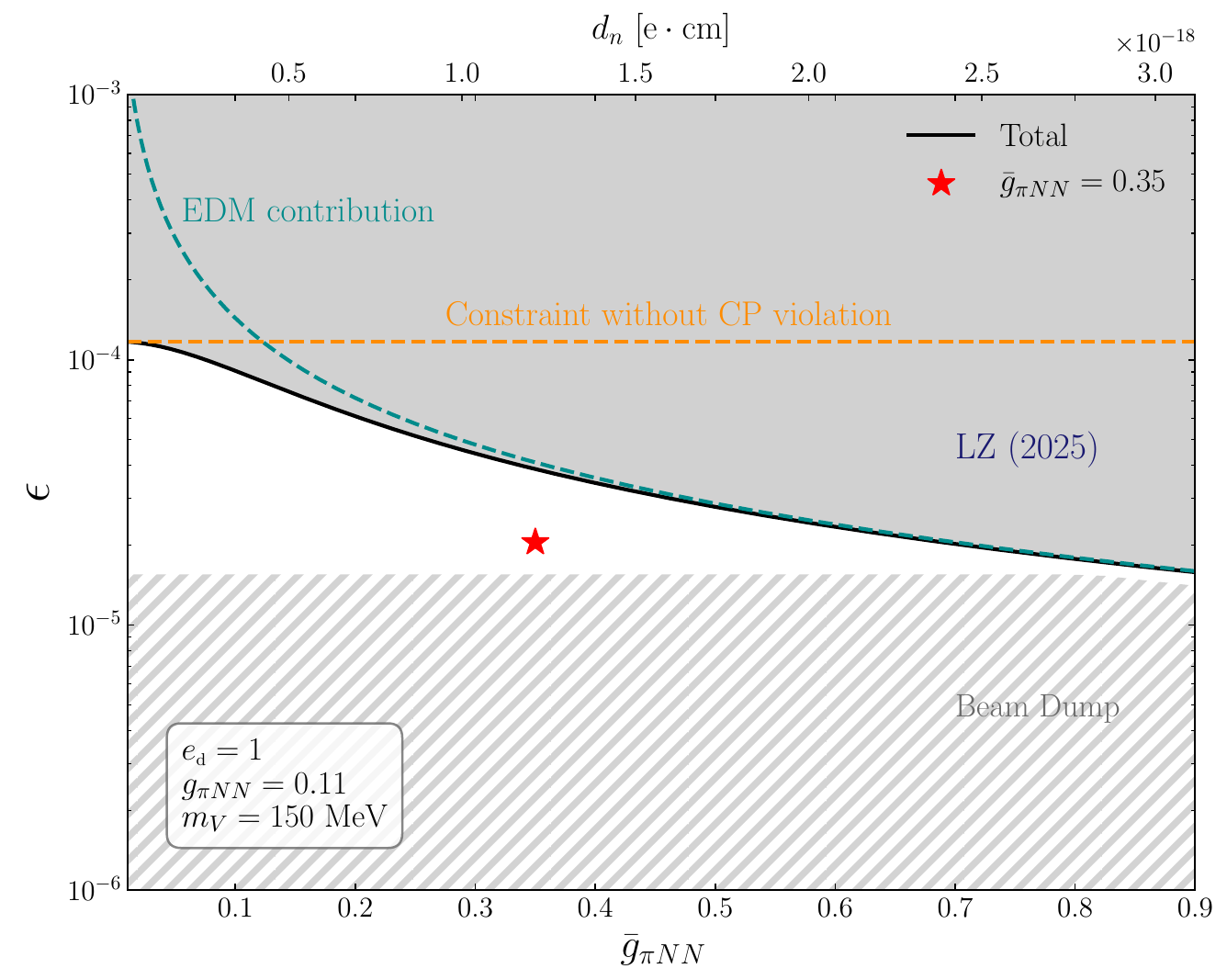}   \includegraphics[width=0.81\linewidth]{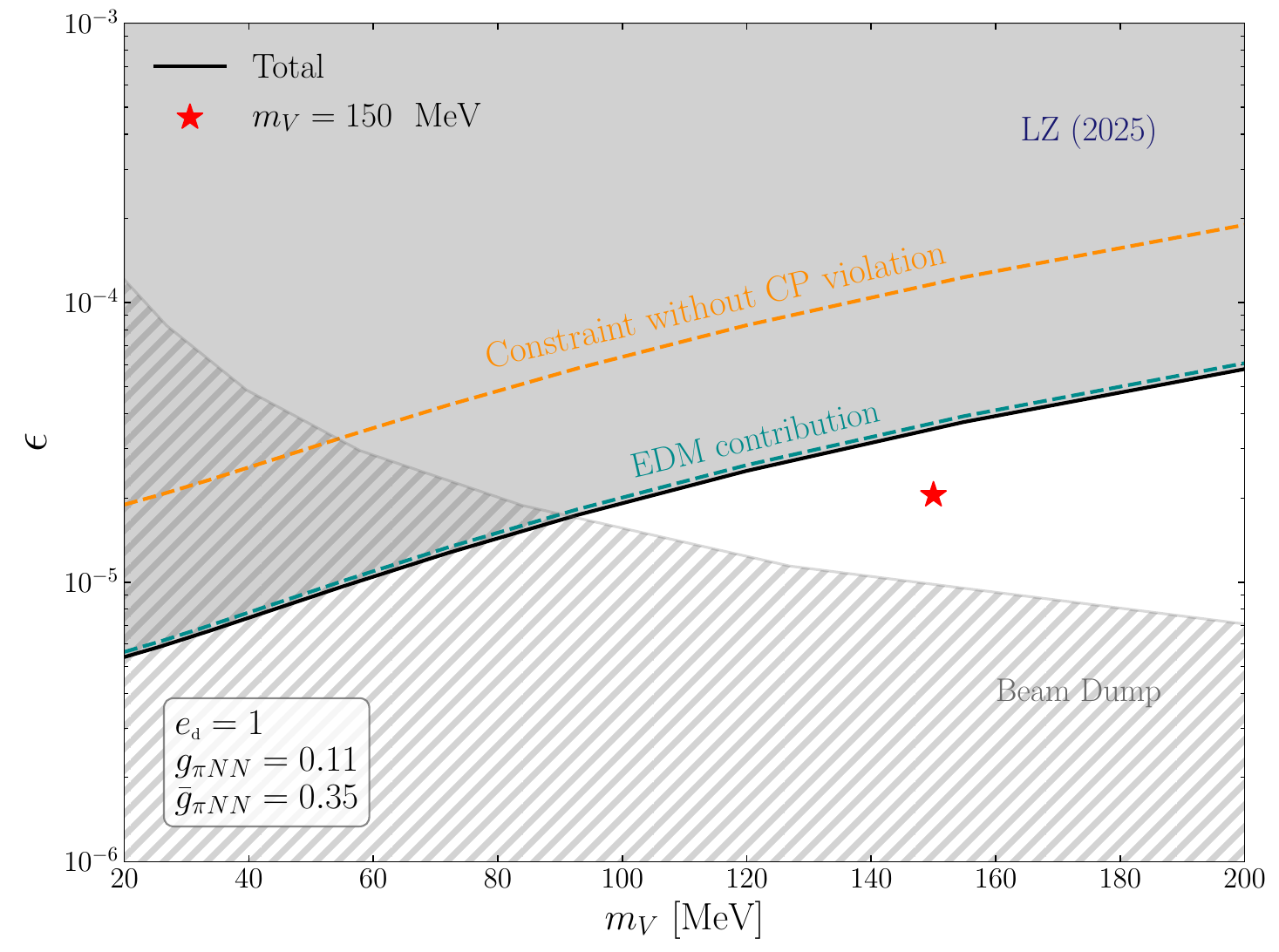} 
    
    \caption{Upper limits at 90$\%$ C.L. in the kinetic mixing $\epsilon$ for the LZ (2025) data release \cite{LZ:2024zvo}. The solid black curve shows the total result including all operators described above, while the orange line shows the CP-conserving limit, obtained by setting $d_n = 0$. \textit{Top panel:} Upper limits in the $(\epsilon, \bar{g}_{\pi NN})$ plane. The upper axis shows the corresponding dark neutron EDM. \textit{Bottom panel}: Upper limits in the $(\epsilon, m_V)$ plane. The beam dump constraints are shown in light hatched gray~\cite{Kyselov:2024dmi, Caputo:2026pdw} .
   }
    \label{fig:limits_gbar}
\end{figure}

We now turn to deriving direct detection limits.  In our analysis we focus on the LZ 2025 data release, as it has been shown to provide the most stringent limits to date on both spin-independent (SI) and spin-dependent (SD) WIMP-nucleon cross-sections for DM masses $\gtrsim 10$ GeV \cite{LZ:2024zvo}.  We use a total number of events analysis. The predicted number of signal events $n_{\rm{sig}}$ is obtained by integrating the recoil spectrum over the corresponding ROI, including the detector efficiency and the total exposure. In our implementation, LZ (2025) is treated as an effective one-bin count experiment following the \textsc{WimPyDD} procedure \cite{Jeong:2021bpl}. If the expected number of signal events can be written as $n_{\rm{sig}}(\lambda) = \lambda \ n_{\rm{sig}}^{\rm{unit}}$, where  $n_{\rm{sig}}^{\rm{unit}} = n_{\rm{sig}}^{\rm{unit}}(\lambda = 1)$, then the 90 $\%$ C.L. upper bound on $\lambda$ can be obtained as, 
\begin{equation}
    \lambda^{\rm{lim}} = \lambda^{\rm{ref}} \frac{N^{90 \%}}{n_{\rm{sig}}(\lambda = \lambda_{\rm{ref}})}
\end{equation}
where $N^{90\%}$ is the 90 $\%$ upper limit on the allowed number of signal events. In the present one-bin recast, it is determined by Poissonian statistics. The corresponding LZ (2025) ROI, exposure, detector efficiency, and number of background events were taken from \cite{LZ:2024zvo}. In our case $\lambda = \epsilon^2$, with all other parameters fixed.

In Fig.~\ref{fig:limits_gbar}, the upper bounds on the kinetic mixing $\epsilon$ are displayed. For small values of $\bar{g}_{\pi NN}$, the limit is set completely by the magnetic dipole and the charge radius contributions --with the coefficients $b_n$ and $\mu_n$  taken as discussed above-- and is therefore independent of the chosen values of the pion-nucleon couplings. As $\bar{g}_{\pi NN}$ increases, the induced dark neutron EDM takes over, and the total limit scales approximately as  $\epsilon_{\textrm{lim}} \propto 1/\bar{g}_{\pi NN}$ in this regime. Comparing our LZ limit with the complementary dark photon beam-dump constraints, we find a viable window in which the CP-violating dark neutron EDM dominates the direct detection signal while remaining compatible with the other bounds. This region is realized for dark neutron masses of the order of $m_n \sim 100$ GeV, pion masses of a few hundred MeV, and pion nucleon couplings of order $\mathcal{O}(0.1)$, for $\bar{g}_{\pi NN}$ somewhat larger than $g_{\pi NN}$. 

\subsection{Annual Modulation}

Our previous results motivate the question of whether direct detection could identify a possible CP-violating origin of the signal. The recoil spectra alone, however, do not provide any distinctive features associated with it. A large differential rate might be interpreted as a hint for an EDM, but it could also be obtained from a larger magnetic moment. To go beyond this rate degeneracy, we exploit the velocity-dependence of the operator $\mathcal{O}_{5}$, which is absent in $\mathcal{O}_{11}$. Annual modulation is a natural observable for this purpose, as it traces the time dependence of the recoil rate induced by the Earth-frame velocity distribution, offering a complementary way of distinguishing direct-detection signals \cite{DelNobile:2015tza, DelNobile:2015rmp, Witte:2016ydc}. This signature has motivated dedicated experimental searches \cite{Freese:2012xd, Froborg:2020tdh, Kahlhoefer:2023}.

\begin{figure}[t!]
    \centering
    \includegraphics[width=0.81\linewidth]{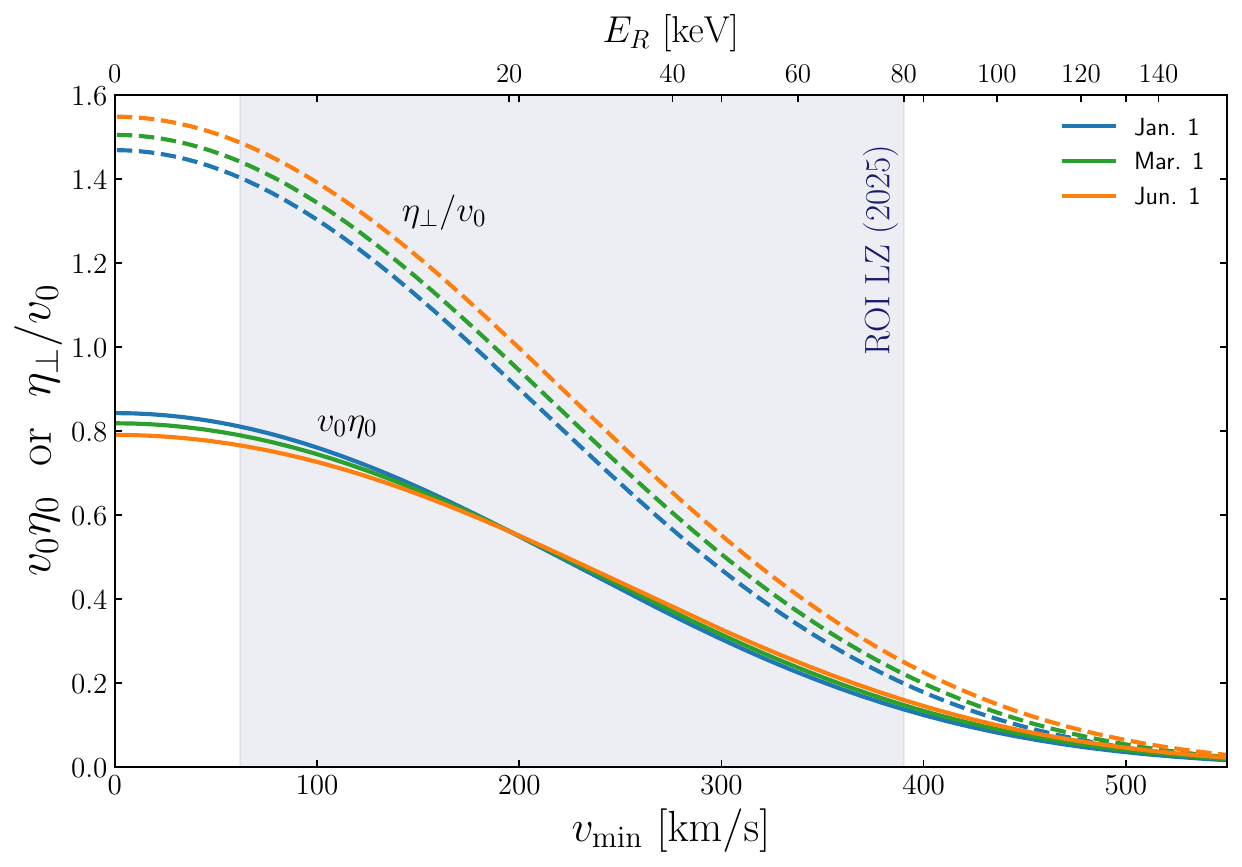}
    \caption{Dimensionless velocity integrals $v_0 \eta_0$ (solid) and $\eta_\perp / v_0$ (dashed) as a function of $v_{\rm{min}}$, evaluated at different times of the year.  The first integral captures the time-dependence associated with the EDM operator $\mathcal{O}_{11}$, while the second one corresponds to the $v_\perp^2$-weighted of the dominant $\mathcal{O}_5$ from the CP-conserving contributions.}
    \label{fig:halo_integrals}
\end{figure}

The origin of this effect can already be understood at the level of the velocity integrals,
\begin{align}\label{halo_integrals}
    &\eta_{0}(v_{\textrm{min}},t) = \int_{v > v_{\textrm{min}}}  \frac{f_{\oplus}(\bm{v},\bm{v}_{E}(t))}{v} d^3 v \ , &&
    &\eta_{\perp}(v_{\textrm{min}},t) = \int_{v > v_{\textrm{min}}} v_{\perp}^{2} \frac{f_{\oplus}(\bm{v},\bm{v}_{E}(t))}{v} d^3 v\,.
\end{align} 
where we adopted a Maxwell-Boltzmann distribution for $f_{\oplus}(\bm{v},\bm{v}_{E}(t))$ with velocity dispersion $v_0$, and $v_{\rm min}$ as defined above. In this notation, the velocity-integral in Eq.~\eqref{rate} reduces to $\eta_0$, for the CP-violating contribution associated with $O_{11}$, and to $\eta_\perp$ for the leading CP-conserving one governed by $O_{5}$. For fixed kinematics, $v_{\perp}$, defined above in terms of $q^2 $, is given by $v_\perp^2 = v^2 - v_{\rm{min}}^2$. 

The DM--nucleus cross section contains target and DM response functions that do not factorize into a single velocity integral \cite{DelNobile:2015rmp}. Nevertheless, the time dependence of the rate enters only through the lab-frame velocity distribution, and the behavior of its corresponding weighted-integrals in Eq.~\eqref{halo_integrals} already provides a useful diagnostic of the modulation pattern. This is illustrated in Fig.~\ref{fig:halo_integrals} for the integrals $v_0 \eta_0$ (color solid) and $\eta_{\perp} / v_{0}$ (color dashed). Following \cite{Hambye:2021xvd}, we multiplied the velocity integrals by appropriate powers of the velocity dispersion $v_{0}$. This normalization makes the plotted quantities dimensionless, so that the two curves can be compared on the same vertical scale without changing their physical content. The different behavior of the two can be understood qualitatively from their velocity weighting. At low $v_{\textrm{min}}$, almost the entire velocity distribution contributes to $\eta_0$. The larger lab speed around June (orange), and the $1/v$ weight makes the integral smaller than in January (blue). At increasing minimum-velocity, the integral is instead controlled by the high-velocity tail, so the June boost increases the number of particles above threshold and $\eta_0$ becomes larger in June. This change of ordering is the usual phase reversal associated with spin-independent interactions $(\mathcal{O}_1)$ within the Standard Halo Model \cite{Lewis:2003bv, Freese:2012xd}. 
By contrast, the factor $v_{\perp}^2$ in $\eta_{\perp}$ suppresses particles close to threshold and gives more weight to the high-velocity part of the distribution. As a result,
the curves at different times remain separated in the same order over most of the $v_{\textrm{min}}$ range. Therefore, for a signal compatible with a dipole interaction, the annual modulation pattern can provide information beyond the time-averaged spectrum. A phase reversal, or a sizeable shift in the time of maximum with respect to the  $\mathcal O_5$-dominated expectation, would be a possible indicator of an EDM CP-violating contribution.

\begin{figure}
    \centering
    \includegraphics[width=0.81\linewidth]{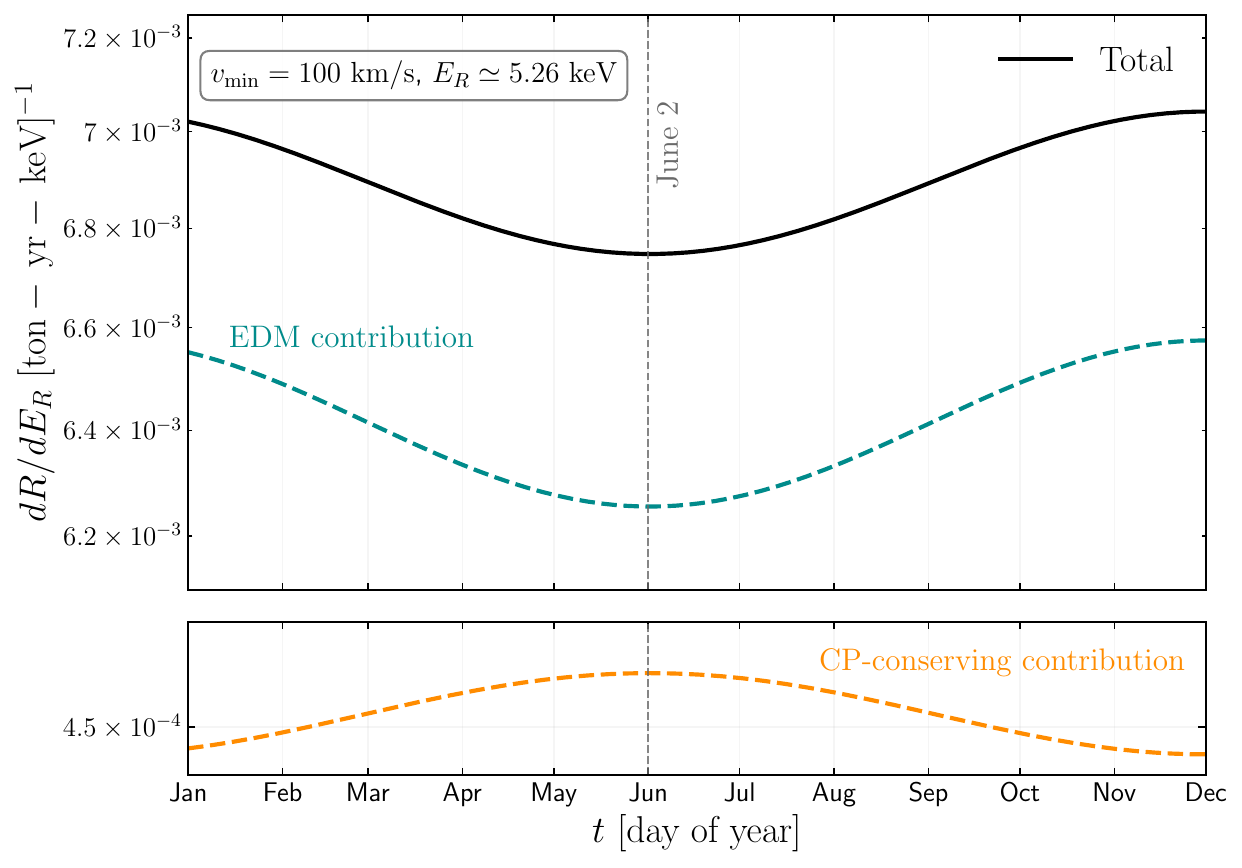}
    \includegraphics[width=0.81\linewidth]{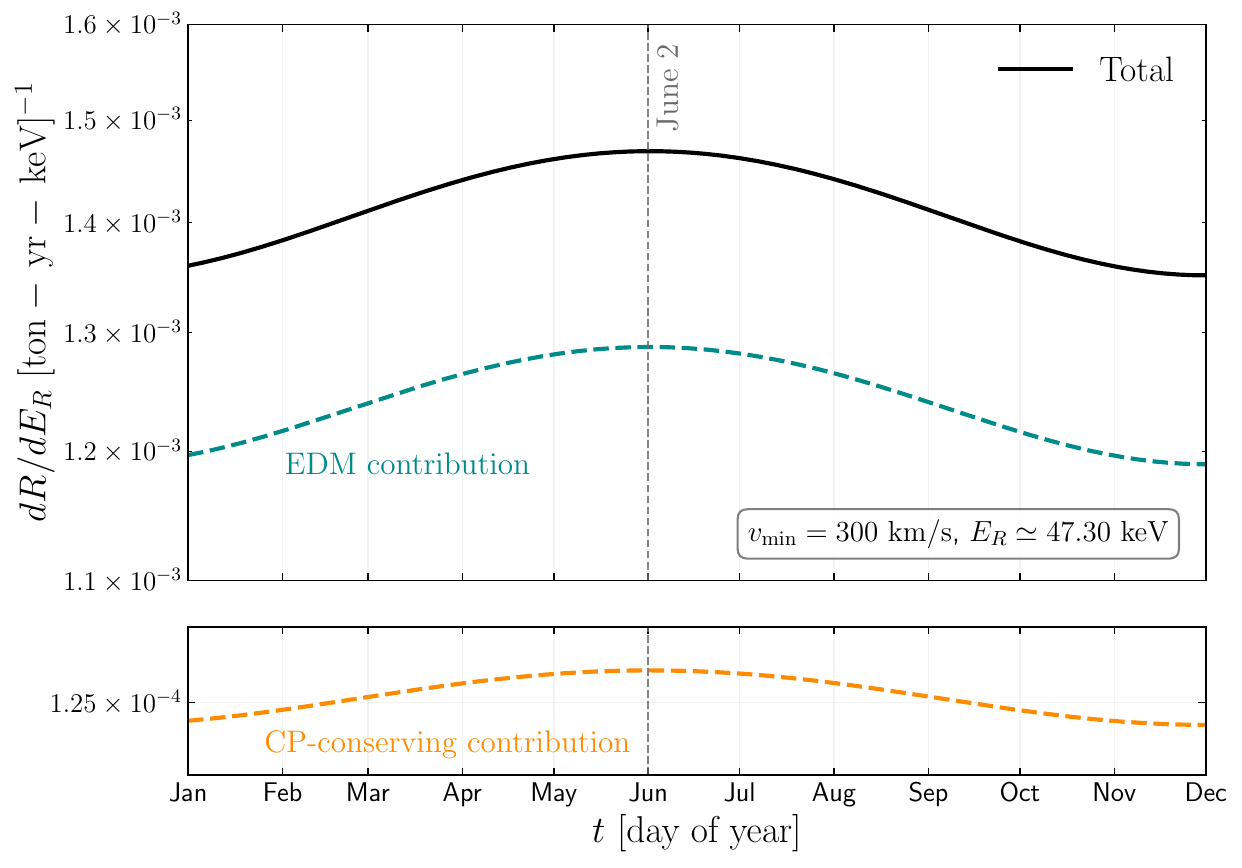}
    \caption{Time dependence of the differential recoil rate, $dR(E_R,t)/ dE_R$, evaluated at two fixed recoil energies, corresponding to $v_{\textrm{min}} = 100$ km/s (\textit{top panel}) and $v_{\textrm{min}} = 300$ km/s (\textit{bottom panel}). In each panel, the upper plots correspond to the total prediction (solid black) together with the leading EDM contribution (dashed cyan), while the lower one shows the CP-conserving contribution (dashed orange). The vertical dashed line marks June $2$. }
    \label{fig:time_dependent_rate}
\end{figure}

We illustrate this point at the level of the differential rate in Fig.~\ref{fig:time_dependent_rate}, where the time dependence of $dR(E_R,t)/dE_R$ is shown at two fixed recoil energies for the benchmark in Table \ref{tab:benchmark-star}. These two choices correspond to $v_{\rm{min}} = 100$ km/s and  $v_{\rm{min}} = 300$ km/s, and therefore probe the two regimes associated with the phase reversal in the velocity-integrals. As expected, in the low-$v_{\textrm{min}}$ regime, the total contribution (solid black) is phase reversed and reaches a minimum close to June 2, while the CP-conserving (dashed orange) remains maximal around that time (top panel). At $v_{\rm{min}} = 300$ km/s the high-velocity tail of the distribution dominates, so both the CP-conserving and the total rate peak close to June 2 (bottom panel). 

This figure should be interpreted as a direct-detection hint of the modulation pattern rather than as an experimental goal. Although a differential rate of $\mathcal{O}(10^{-3}- 10^{-4})$ events/(ton-year-keV) could lead to order one-event counts in the next generation of Xenon detectors \cite{Baudis:2024jnk}, the January-June separation (sometimes called the \textit{modulation fraction}) is only of the $(1 - 3)\%$ level, which would imply that (even with an ideal detector) resolving a one-event seasonal difference might require roughly 100 times more signal events than just the time-averaged rate itself. If such a level of sensitivity were ever achieved, the experiment would be already in the regime where atmospheric neutrinos become detectable, providing an irreducible background for DM searches at $\mathcal{O}(100)$ GeV \cite{OHare:2021utq}. In that case, the time-variation of this neutrino background \cite{Zhuang:2024exm} would have to be included before annual modulation could be used as a possible discriminator between dipole interactions. In this respect, the recent indication of solar $^8B$ neutrinos in large volume Xe detectors already illustrates the challenges of disentangling neutrino signals from intrinsic backgrounds and possible light WIMPs \cite{PandaX:2024muv, XENON:2024ijk, LZ:2025igz}.

\section{Putting Everything Together: A Benchmark Point}\label{sec:bench}

We now collect the results of the previous sections into a representative benchmark point, reported in Table~\ref{tab:benchmark-star}, which simultaneously realizes the desired DM self-interactions and satisfies current cosmological, astrophysical, and laboratory constraints. In this benchmark, DM consists almost entirely of dark neutrons, while dark protons and charged dark pions constitute only negligible fractions of the total DM abundance. The benchmark corresponds to the red star in Fig.~\ref{fig:sidm_plot}. Since DM is neutral and asymmetric, halo dynamics is governed almost entirely by $nn$ scattering mediated by the neutral pion (see Sec.~\ref{sec:sidm}). The viscosity cross section, shown in the inset of Fig.~\ref{fig:sidm_plot}, reaches large values at low velocities while remaining compatible with constraints from galaxy clusters at higher velocities, thereby realizing the desired SIDM phenomenology.

DM interactions with SM matter proceed through the loop-induced electromagnetic form factors discussed in Sec.~\ref{sec:dd}. The corresponding parameter space is shown in Fig.~\ref{fig:limits_gbar}. Direct-detection constraints, dominated by the CP-violating electric dipole moment induced by the topological $\theta$ angle, together with the limits on the dark photon mass and kinetic mixing from beam-dump experiments, leave an allowed region of parameter space with $m_V\lesssim m_{\pi^0}$. The benchmark point lies within this region and is indicated with a red star. Its predicted differential recoil spectrum is shown in Fig.~\ref{fig:rates}. The benchmark has been chosen to illustrate the simultaneous realization of the desired SIDM and direct-detection phenomenology. It is located close to the peak in Fig.~\ref{fig:sidm_plot}, where the overlap between the two requirements is favorable. A systematic exploration of the parameter space and the degree of tuning associated with this overlap lies beyond the scope of the present work. 

Finally, the benchmark is consistent with the cosmological constraints. The kinetic mixing is sufficiently large to establish thermal equilibrium between the dark sector and the SM thermal bath in the early universe, while the DM mass lies in the appropriate range for asymmetric DM. Dark protons efficiently convert into dark neutrons, leaving a relative abundance $Y_p/Y_n\lesssim10^{-5}$, while the abundance of charged dark pions is reduced to less than $10^{-4}$ of the total DM abundance after annihilation into neutral dark pions and dark photons. The neutral dark pions subsequently decay as $\pi^0\to VV$, followed by the decay of the dark photons into $e^+e^-$ with a lifetime much shorter than one second, thereby safely evading the constraints from Big Bang Nucleosynthesis.  

Before concluding, let us remark that our benchmark should be compared with previous studies of self-interacting DM mediated by light scalar particles~\cite{Kahlhoefer:2017umn}. In those scenarios the mediator is introduced as an elementary degree of freedom whose couplings are responsible both for DM self-scattering and for direct-detection signals, while the DM abundance is typically determined through thermal freeze-out. In the present framework, by contrast, DM is composite and naturally accommodates an asymmetric origin, which is particularly well motivated in the presence of CP-violating dynamics. Furthermore, the phenomena of self-interactions and direct detection arise from different low-energy degrees of freedom. Self-scattering is controlled by dark-pion exchange, whereas direct detection proceeds through the dark-photon portal and the electromagnetic structure of the dark baryon. This separation leads to a phenomenological picture that differs from simplified spin-0 mediator models and avoids some of the tensions emphasized in previous analyses, see e.g.~\cite{Bernal:2015ova}.

\section{Conclusions and outlook}
\label{sec:conclusions}

In this work, we have explored the phenomenology of baryonic DM in QCD-like theories with a non-vanishing topological angle. The key observation is that the pseudo-Goldstone bosons generically present in these theories provide naturally light mediators of DM interactions. In the absence of CP violation, their effects on dark-baryon scattering are typically suppressed. A physical $\theta$ angle qualitatively changes this picture, turning the pseudo-Goldstone bosons into mediators of an attractive long-range force between DM particles and leading to sizable self-interactions in astrophysical halos of DM.

To illustrate these ideas, we analyzed in detail a simple QCD-like realization inspired by the low-energy dynamics of ordinary QCD. We computed the DM self-scattering cross section arising from CP-violating pion exchange and showed that it can reach values relevant for the evolution of DM halos while remaining compatible with constraints from galaxy clusters. We further demonstrated that the resulting interactions exhibit the characteristic velocity dependence expected from long-range forces, with the strength of the self-scattering controlled by the topological angle. These results provide an explicit realization of the SIDM paradigm within a confining dark sector, where both the light mediator and its CP-violating couplings arise naturally from the underlying strong dynamics.

In addition, we performed a detailed study of the direct-detection signatures arising in the presence of a kinetically mixed dark photon. We computed the electromagnetic form factors of the dark baryon and quantified the impact of the electric dipole moment induced by the $\theta$ term relative to the charge-radius and magnetic-dipole interactions that are already present in the CP-conserving limit and whose magnitude is set by the underlying strong dynamics.  After matching these interactions onto the non-relativistic effective theory relevant for nuclear recoils, we studied the resulting recoil spectra, current experimental constraints, and the extent to which CP-violating contributions can be distinguished from CP-conserving interactions through their distinctive operator structure and annual modulation signatures. We found that the CP-violating electric dipole moment can dominate the scattering rate over sizeable regions of parameter space, significantly enhancing the experimental sensitivity to the dark sector.  As a result, the same topological angle that governs the DM self-interactions can also determine the strength of the direct-detection signal, providing a direct link between halo-scale and laboratory observables.  Although the different operator structures lead to distinct annual-modulation patterns, we found that experimentally resolving these differences would require roughly two orders of magnitude more signal events than the detection of the time-averaged rate, making such discrimination  challenging in practice.

Although our quantitative analysis focused on a specific QCD-inspired realization, the main lesson extends well beyond this example. The ingredients underlying our results—a physical $\theta$ angle, stable baryonic DM, and light pseudo-Goldstone bosons—can arise naturally in many confining theories. In such theories, the topological angle can simultaneously influence the long-range interactions of DM and its couplings to ordinary matter through electric dipole moments. The framework studied here therefore illustrates a broader possibility: that topological CP violation may serve as a common origin for both halo-scale and laboratory signatures of composite DM.

Future studies of self-interacting DM and increasingly sensitive direct-detection experiments may provide a  window into the non-perturbative dynamics of composite dark sectors and the possible existence of a dark topological angle. Furthermore, the presence of the CP-violating interactions induced by the topological $\theta$ angle may provide additional motivation for an asymmetric origin of the DM abundance. This scenario motivates further investigation of viable mechanisms for generating the dark asymmetry, possibly in connection with the baryon asymmetry of the visible sector. Interestingly, the CP violation induced by the topological $\theta$ angle may itself play a relevant role in this context. A quantitative study of the resulting relic abundance within such an asymmetric framework, and its connection to the underlying strong dynamics, is left for future work.

\section*{Acknowledgments}
We are indebted to Andrew Cheek, Yi Chung, Juan Herrero García, Manoj Kaplinghat, Helena Kolesova, Nuria Rius, Óscar Zapata, and Vanessa Zema for valuable discussions. G.L. is supported by the INFN Cabibbo Fellowship, call 2025.
C.G.C. and P.F. are  supported by a Ramón y Cajal contract with Ref.~RYC2020-029248-I, the Spanish National Grant PID2022-137268NA-C55 and Generalitat Valenciana through the grant CIPROM/22/69. P.F. is also supported through the GenT Excellence Program (CIESGT2024-007).
\appendix

\newpage
\section{Chiral interactions for \texorpdfstring{$N_f = 2$}{Nf = 2}}\label{app:chiralLag}

The low-energy dynamics of dark baryons, dark pions and dark photons below the confinement scale $\Lambda$ is described by chiral perturbation theory in analogy with ordinary QCD~\cite{Pich:1995bw,Scherer:2005ri}. For simplicity, the term \textit{dark} will be omitted, while still referring to dark states throughout. First, we review the relevant interactions for the benchmark model of Sec.~\ref{sec:sec2}, with $N_c=3$ and $N_f=2$ and for simplicity take degenerate flavors of light quarks of mass $m_1=m_2\equiv m$. In particular, we focus on the trilinear baryon-pion couplings of Eq.~\eqref{eq:baryonPionLag} and the couplings of baryons and pions with photons, relevant for the computation of the electric dipole moment in Sec.~\ref{sec:dd} and the lifetime of the neutral pion.

In a model with $N_f=2$ light flavors, baryons and pions are described in terms of the fields $N$ and $\bm{\pi}$ as defined in Eq.~\eqref{eq:doublettriplet}. The dark photon is introduced as a vector current.
At the leading order the interactions are described by~\cite{Pich:1995bw,Scherer:2002tk}
\begin{equation}\label{pich_lagrangian}
\begin{split}
    \mathcal{L}_{ N\pi V}^{(N_f=2)}&=e_{\rm d}\overline{N}\gamma_\mu V^\mu Q_N N-g_A\overline{N}\gamma_\mu\gamma_5\left( \frac{\partial^\mu \bm{\pi}\cdot\bm{\sigma}}{2f_\pi}\right)N-b_1\left(\overline{N}\chi_+^{(2)} N\right)\\
   &-{ie_{\rm d}} V_\mu(\partial^\mu{\pi}^+\pi^--\partial^\mu{\pi}^-\pi^+)-(e_{{d}}^2 N_c/48\pi^2 f_\pi)\pi^0V_{\mu\nu}\widetilde{V}^{\mu\nu} +\cdots\,,
\end{split}
\end{equation}
where $Q_N={\rm diag}(1,0)$ is the nucleon charge matrix, $b_1$ is a coupling constant with inverse mass dimension, while $\cdots$ represent additional term for processes not relevant for this work.  The dependence on the quark mass $m$ and the topological angle $\theta$ is contained in $\chi_+^{(2)}$, defined as
\begin{equation}
\label{eq:chip0}
    \chi_+^{(2)}= 
    e^{-\frac{i\bm{\pi}\cdot\bm{\sigma}}{2f_\pi}}
    \chi^{(2)}
    e^{-\frac{i\bm{\pi}\cdot\bm{\sigma}}{2f_\pi}}
    +h.c.\,,\qquad\chi^{(2)}=m_\pi^2[1+i\tan(\theta/2)]\,.
\end{equation}

\paragraph{Baryon-pion interactions.} Expanding the Lagrangian and integrating by parts the derivative term,
one recovers the effective interaction of Eq.~\eqref{eq:baryonPionLag} with
\begin{equation}
     \mathcal{L}^{(\pi NN)} = \underbrace{\frac{g_{A}}{f_{\pi}}m_{N}}_{ \equiv \ g_{\pi NN}} \hspace{0.65mm} \bar{N} i \gamma_{5} \bm{\sigma} \cdot \bm{\pi} N - \underbrace{2 b_{1}\tan(\theta/2) \left( \frac{m_{\pi}^{2}}{f_{\pi}} \right) \hspace{0.5mm}}_{\equiv \ \bar{g}_{\pi NN}} \bar{N} \bm{\sigma} \cdot \bm{\pi} N\,.
\end{equation}
The first relation is the Goldberger-Treiman relation for the present model.
As in \cite{Antipin:2015xia}, the $\theta$-dependent mass terms generate the CP-violating pion-nucleon vertex, while the ordinary CP-conserving one follows from the derivative axial coupling. Notice that the perturbative expansion in the couplings works as long as $g_{\pi NN},\bar{g}_{\pi NN}\lesssim\sqrt{4\pi}$. Using $N_c=3$ and $f_\pi\sim \sqrt{N_c}m_n/4\pi$ , this implies the conditions $g_A\lesssim 0.5$ and $(b_1\,\text{GeV})\,\tan(\theta/2)\lesssim 100\,(m_n/100\text{ GeV})(500\text{ MeV}/m_\pi)^2 $.

\paragraph{Photon-pion interactions.}
The second line of Eq.~\eqref{pich_lagrangian} describes the  photon-pion interactions relevant to our computations. The first term involves the charged pions and enters the computation of the loop diagram on the right in Fig.~\ref{fig:dark-neutron-em-vertex}. The second term involves the neutral pion and is present whenever the charges of the quarks are anomalous under the  $U(1)_{\rm{d}}$ interactions. Analogously to ordinary QCD, the coefficient of the anomalous term is proportional to ${\rm Tr}[Q_{\rm d}^2\sigma_3]=1/3$.

\paragraph{Baryon-photon interactions.}
Expanding the nucleon field, the first term of Eq.~\eqref{pich_lagrangian} reduces to $e_{\rm d}\,\bar{p}\gamma_\mu V^\mu p$. As expected, the photon interacts at tree level with the proton and not with the neutron.

\section{Chiral interactions beyond \texorpdfstring{$N_f=2$}{Nf=2}}
\label{sec:UVmodel}

The benchmark model considered thus far, motivated by its analogy with the SM, constitutes a particular example within a broader class of confining theories. In this section, we turn to the generic features of QCD-like theories with gauge group $SU(N_c)$ and $N_f$ flavors of light Dirac quarks transforming in the fundamental representation, described by the Lagrangian in Eq.~\eqref{eq:L}.
This is invariant under a global $U(1)_V$ symmetry, corresponding to the simultaneous transformation of left and right quarks $q_{L,R}\to e^{i\alpha}q_{L,R}$, which is the analogue of baryon number of the SM. Furthermore, if $m_q\ll \Lambda$, this Lagrangian exhibits an approximate $SU(N_f)_L \times SU(N_f)_R$ symmetry, corresponding to independent chiral transformations of the left- and right-handed fermion components in flavor space. 
In particular, a generic axial transformation takes the form
\begin{align}
q_L \rightarrow e^{-i\theta Q/2} q_L\,, &&  \text{and}&&q_R \rightarrow e^{i\theta Q/2} q_R \,,
\end{align}
where $Q$ is an arbitrary $N_f\times N_f$ hermitian matrix. Being anomalous under the $SU(N_c)$ gauge interactions, such a transformation induces not only a change in the mass matrix but also a shift in the $\theta$ angle. Following closely the derivation of Ref.~\cite{Garcia-Cely:2025flv}, we obtain
\begin{align}
\theta \to \theta(1 - \text{Tr}\,Q)\,, 
&&
\text{and}
&&
M \to M_\theta= e^{i\theta Q/2} M e^{i\theta Q/2}\,.
\end{align}
Hence, one can effectively move the $\theta$ parameter from the $F\tilde{F}$ term into the mass matrix by imposing 
\begin{equation}
\text{Tr}\,Q = 1   \,.
\label{eq:traceCondition}
\end{equation}
In particular, one may take $Q$ diagonal such that $Q=\text{diag}(q_1,...,q_{N_f}) $.

\subsection{Effective low-energy Lagrangian for mesons}

Below the scale of gauge confinement $\Lambda$, the quarks
hadronize into gauge-invariant bound states: baryons, composed of $N_c$ quarks combined antisymmetrically in color, and mesons, which are $\bar q q$ states.
It is believed that the fermion condensate $\langle{\bar{q}q\rangle}$ spontaneously breaks the chiral symmetry to its diagonal subgroup, $SU(N_f)_L \times SU(N_f)_R \rightarrow SU(N_f)_V$. This leads to the emergence of $N_f^2 - 1$ pseudo-Goldstone bosons, $\pi^a$, which we denote generically as pions, whose dynamics at sufficiently low energies is described in chiral perturbation theory~\cite{Pich:1991fq,Scherer:2002tk} by  
\begin{align}\label{eq:chiralLag}
	\mathcal{L}_{\rm eff} = \frac{f_\pi^2}{4} \text{Tr}[\partial_\mu U^\dagger \partial^\mu U] + \frac{f_\pi^2}{2} B_0 \text{Tr}[M_\theta^\dagger U+ U^\dagger M_\theta]\,+\mathcal{L}_{\rm WZW}\,, &&
    \text{with}
    && U =  e^\frac{i\pi^a\lambda^a}{f_\pi} \,.
\end{align}
We normalize the generators of $SU(N_f)$ such that $\operatorname{tr}[\lambda^a \lambda^b] = 2\delta^{ab}$, and parametrize the fermion condensate in terms of the dark meson constant, $f_\pi$, as $\langle \bar{q} q \rangle = -B_0 f_\pi^2$. Note that it is  expected that the confinement scale is of the order $\Lambda\sim 4\pi f_\pi/\sqrt{N_c}$~\cite{Kamada:2022zwb}. The last term of Eq.~\eqref{eq:chiralLag} is the usual Wess-Zumino-Witten interaction~\cite{Wess:1971yu,Witten:1983tw}.

Let us emphasize a few points. The broken generators belong to the coset $(SU(N_f)_L\times SU(N_f)_R)/SU(N_f)_V$, corresponding to the chiral transformations $q_{L,R}\to \exp[\mp i\alpha^a\lambda^a]q_{L,R}$, and they are in a one-to-one correspondence with the light pseudo-scalar mesons $\pi_a$. As a result of the axial nature of the broken generators, the pions are pseudo-scalar particles. For $\theta=0$, 
their interactions preserve CP symmetry, so that in that limit $(i)$ with the exception of anomalous topological terms~\cite{Wess:1971yu,Witten:1983tw}, the pion self-interactions, given in Eq.~\eqref{eq:chiralLag}, contain an even number of fields. ($ii$) The interaction between one pion and two (fermionic) baryon fields, denoted as $N$, is of the form $\sim \bar{N}\gamma_5\bm{\sigma}\!\cdot\!\bm{\pi}\,N$, corresponding to the ${g}_{\pi NN}$ term in Eq.~\eqref{eq:baryonPionLag}, where the $\gamma_5$ insertion is needed for CP conservation. On the contrary, the $\bar{g}_{\pi NN}$ coupling only arises in the presence of a non-vanishing $\theta$ angle, as we discuss in the next section. These results are quite general. Indeed, Vafa and Witten showed~\cite{Vafa:1983tf} that in QCD-like theories with fundamental fermions transforming in vector-like representations of the gauge group and vanishing $\theta$ angle, only axial symmetries can undergo spontaneous breaking. As a consequence, the associated Goldstone bosons are pseudo-scalars, and their interactions are constrained as described above. The presence of a nonzero $\theta$ angle modifies this picture.

Finally, we note that
the $SU(N_f)_V$ symmetry is an exact symmetry in the limit of degenerate quarks and no additional external source of breaking. 

As shown in Ref.~\cite{Garcia-Cely:2025flv}, the matrix $Q$ must satisfy the condition $\sin(\theta Q)=\alpha(\theta)M^{-1}$, with $\alpha(\theta)$ being some function canceling the linear terms in the pion fields in the effective Lagrangian (or, in other words, selecting the correct vacuum). This leads to the general structure of the $\theta$-dependent quark mass matrix

\begin{equation}
    M_\theta=\cos(\theta Q)M+i\alpha(\theta)\mathbb{1}\,.
\end{equation}

The general expressions for $Q$ and $\alpha(\theta)$ have been derived in Ref.~\cite{Garcia-Cely:2025flv}. Here, we only report the results in two simple limits:

\begin{itemize}
    \item for degenerate quarks $m_i=m$, $\alpha(\theta)=m\sin(\theta/N_f)$ and $Q=\mathbb{1}/N_f$.
    \item For $\theta\ll 1$, $\alpha(\theta)=\theta/{\rm Tr}M^{-1}$ and $Q=M^{-1}/{\rm Tr}M^{-1}$.
\end{itemize}

Finally, the interactions among mesons and photons can be obtained as follows: first, we promote the derivative to the corresponding covariant version, $D^\mu U=\partial^\mu U- i e_{\rm{d}} V_\mu [{Q_{\rm d}},{U}]$. Second, we extend the WZW terms to include the gauge field. While the complete set of operators can be found in~\cite{Witten:1983tw}, the most relevant for our discussion are of the form $\pi^a V_{\mu\nu}\tilde{V}^{\mu\nu}$, which give rise to the anomalous decay of the neutral mesons. The corresponding coefficient is proportional to ${\rm Tr}[Q_{\rm d}^2\lambda^a]$, where $\lambda^a$ is the broken axial generator associated with $\pi^a$. 

\subsection{Meson couplings to the lightest baryon for arbitrary flavors}

The CP-violating term in Eq.~\eqref{eq:baryonPionLag} allowed us to derive the
Yukawa potential between baryons for $SU(2)$ flavor symmetry. We now derive the
corresponding CP-odd meson--baryon coupling for arbitrary $N_f$, showing that
the Yukawa potential induced by the $\theta$-dependent CP-violating term is a
generic feature of a larger class of theories.

Baryons are gauge-invariant bound states made of $N_c$ quarks
anti-symmetrized in color. For odd $N_c$, these states are fermions. In the
following we focus on $N_c=3$, while keeping the number of flavors arbitrary,
$N_f\geq2$.\footnote{For $N_f=1$ there is no $SU(N_f)_A$ and the axial
$U(1)_A$ is anomalous, resulting in a would-be Goldstone boson with a large
mass $m\sim \Lambda/\sqrt{N_c}$, similar to the $\eta'$ of the SM,
with all other mesons heavier or comparable. On the other hand, also for only one flavor, the lightest
dark baryon has spin $N_c/2$; see for
instance~\cite{Morrison:2020yeg,Berbig:2024uwm} for models of higher-spin
baryon DM. }
Then, the baryon is made of three quarks and the color wavefunction is
completely anti-symmetric. Fermi statistics then implies that the remaining
spin-flavor wavefunction must be symmetric. To identify the flavor
wavefunction, we first consider the spin part. The lightest baryon is expected
to be the spin-$1/2$ state, obtained by first combining two quark spins into a
symmetric spin-one pair and then combining this pair with the third spin-$1/2$
quark to form total spin $1/2$. This spin wavefunction has mixed permutation
symmetry: it is symmetric under the exchange of the two quarks in the spin-one
pair, but it is not invariant under arbitrary permutations of the three
quarks. Consequently, the flavor wavefunction must carry the corresponding
mixed symmetry, so that the product of the spin and flavor parts contains the
fully symmetric spin-flavor wavefunction required by Fermi statistics.

The corresponding flavor representation is described by the Young tableau with
two boxes in the first row and one box in the second
row~\cite{Antipin:2015xia}. We denote the associated
baryon field by $N^{ij;\ell}$, where the semicolon separates the two rows of
the Young tableau. The tensor is symmetric in the indices belonging to the
first row and obeys the irreducibility constraint
\begin{equation}
    N^{ij;\ell}=N^{ji;\ell}\,,
    \qquad
    N^{ij;\ell}+N^{j\ell;i}+N^{\ell i;j}=0\,.
    \label{eq:N_tensor_constraint}
\end{equation}

Following the standard practice for unitary groups, the conjugate field is denoted by $\bar N_{ij;\ell}$ (which satisfies analogous constraints). Let us discuss special values of $N_f$.
\begin{itemize}
\item For $N_f=2$, one can use the invariant tensor $\epsilon_{ij}$
to define
\begin{equation}
    N^i=\epsilon_{j\ell}N^{ij;\ell}\,.
\end{equation}
The constraint in Eq.~\eqref{eq:N_tensor_constraint} implies that this
contains all independent components of $N^{ij;\ell}$. Thus, for $N_f=2$, the
lightest baryon transforms as a doublet of $SU(2)_V$, in agreement with Eq.~\eqref{eq:doublettriplet}, and the treatment of the main text. For degenerate quark masses, one recovers the results of Appendix~\ref{app:chiralLag}.\

\item For $N_f=3$, one can use the invariant tensor $\epsilon_{ijk}$ to define
\begin{equation}
    N^i{}_j=\epsilon_{jab}N^{ia;b}\,.
\end{equation}
The constraint in Eq.~\eqref{eq:N_tensor_constraint} implies that this matrix
is traceless,
\begin{equation}
    N^i{}_i=0\,.
\end{equation}
Hence, for $N_f=3$, the lightest baryon transforms as the adjoint
representation of $SU(3)_V$, namely an octet. 
\item For $N_f>3$, the epsilon tensor
does not reduce the mixed tensor to a matrix field, and the natural
description remains in terms of $N^{ij;\ell}$.
\end{itemize}

We now write the leading non-derivative interactions between the baryons and
the pseudo-Goldstone bosons. These are obtained by coupling the baryon tensor
to the field $\chi_+$ 
\begin{equation}
    \chi_+
    =
    e^{-\frac{i\pi^a\lambda^a}{2f_\pi}}
    \chi
    e^{-\frac{i\pi^a\lambda^a}{2f_\pi}}
    +h.c.\,,
    \qquad
    \chi=2B_0M_\theta\,.
    \label{eq:chidef}
\end{equation}
which generalizes Eq.~\eqref{eq:chip0}. 
There are two
independent contractions,
\begin{equation}
    \mathcal{L}_{N\pi}
    =
    -b_1\,\bar N_{im;\ell}(\chi_+)^m{}_n N^{in;\ell}
    -b_2\,\bar N_{ij;m}(\chi_+)^m{}_n N^{ij;n}
    +\cdots\,,
    \label{eq:LNpi_tensor}
\end{equation}
where $b_{1,2}$ are unknown low-energy constants. The  additional terms involve derivatives of the pion field.  Focusing on the CP-odd part, one obtains
\begin{equation}
    \mathcal{L}_{N\pi}^{\rm CP\text{-}odd}
    =
    -2iB_0\alpha(\theta)
    \left[
    b_1\,\bar N_{im;\ell}(U^\dagger-U)^m{}_nN^{in;\ell}
    +
    b_2\,\bar N_{ij;m}(U^\dagger-U)^m{}_nN^{ij;n}
    \right]\,.
    \label{eq:LNpi_CPodd_tensor}
\end{equation}
Expanding to leading order in the pion field gives the CP-odd Yukawa
interaction
\begin{equation}
    \mathcal{L}_{N\pi}^{\rm CP\text{-}odd}
    =
    -\frac{4B_0\alpha(\theta)}{f_\pi}\,
    \pi^a
    \left[
    b_1\,\bar N_{im;\ell}(\lambda^a)^m{}_nN^{in;\ell}
    +
    b_2\,\bar N_{ij;m}(\lambda^a)^m{}_nN^{ij;n}
    \right]\,.
    \label{eq:CPodd_Yukawa_tensor}
\end{equation}
The corresponding Yukawa potential is straightforwardly obtained from this expression
(see, e.g., Ref.~\cite{Petraki:2016cnz}). Notice that the previous equation
applies for generic values of $\theta$, which may be an $\mathcal{O}(1)$
parameter. For $N_f=3$, using $N^i{}_j=\epsilon_{jab}N^{ia;b}$, this expression can be
rewritten in the usual octet notation.
The derivative  baryon-meson interactions and the couplings of baryons to photons can be obtained by generalizing the Lagrangian in Eq.~\eqref{pich_lagrangian} to an arbitrary number of flavors and to generic quark masses and charges.

\section{Alternatives to asymmetric DM}
\label{alternatives}

In the main text, we introduced a dark photon portal between the dark sector and the SM. On the one hand, this provides the phenomenology for DM direct detection searches discussed in Sec.~\ref{sec:dd}. On the other hand, the presence of a light dark photon allows the decay of the dark neutral pions, whose energy density would otherwise overclose the universe. Additionally, we focused on the asymmetric DM scenario, in which the DM cosmological relic abundance is determined by a primordial asymmetry. In this Appendix, we briefly comment on alternative cosmological scenarios.

First, keeping the dark photon portal option, the DM relic abundance could be obtained through the usual symmetric thermal freeze-out. In such a case, dark neutron pairs annihilate into dark pions, freezing-out at $T\sim m_n/20$ and leaving a constant relic. Strong dynamics typically selects DM masses of the order of $1-100$ TeV, although the uncertainties are large. The heavier dark protons convert to the lighter dark neutrons as discussed in the main text. The relevant processes for indirect detection searches  are DM annihilations which involve the exchange of either dark photons or dark pions. In the former case, given the neutrality of DM, the corresponding cross sections are suppressed by loop factors. In the latter case, the cross sections are velocity-suppressed in some region of the parameter space (indeed the dominant contribution to the cross section for DM annihilations into neutral dark pions is $p$-wave suppressed as long as ${g}_{\pi NN}\ll \bar{g}_{\pi NN}$). While plausible, this scenario appears to be less favorable than the asymmetric alternative, particularly in light of the large DM masses expected from thermal production.

\begin{figure}[t!]
    \centering
    \includegraphics[width=0.81\linewidth]{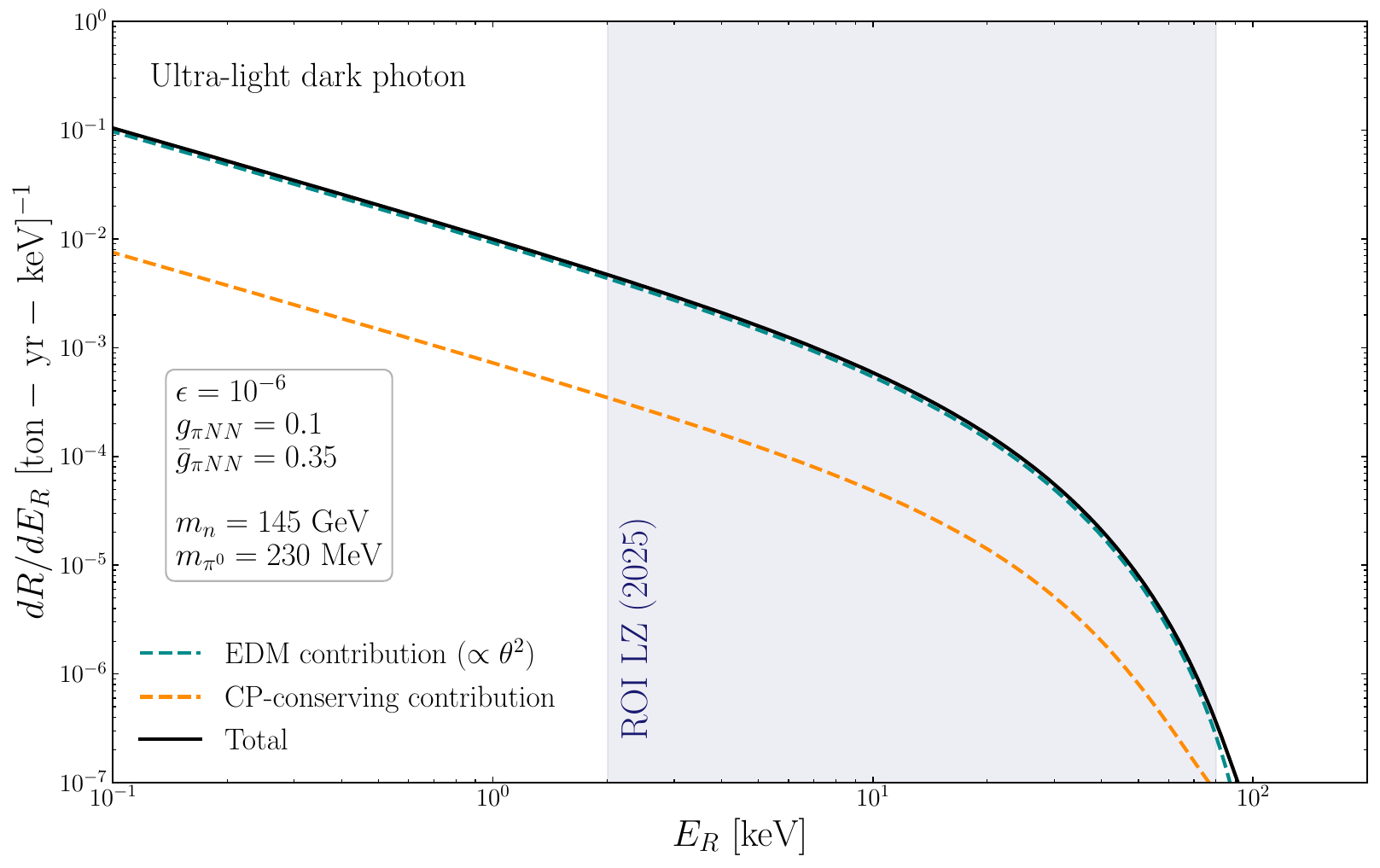}
    \includegraphics[width=0.83\linewidth]{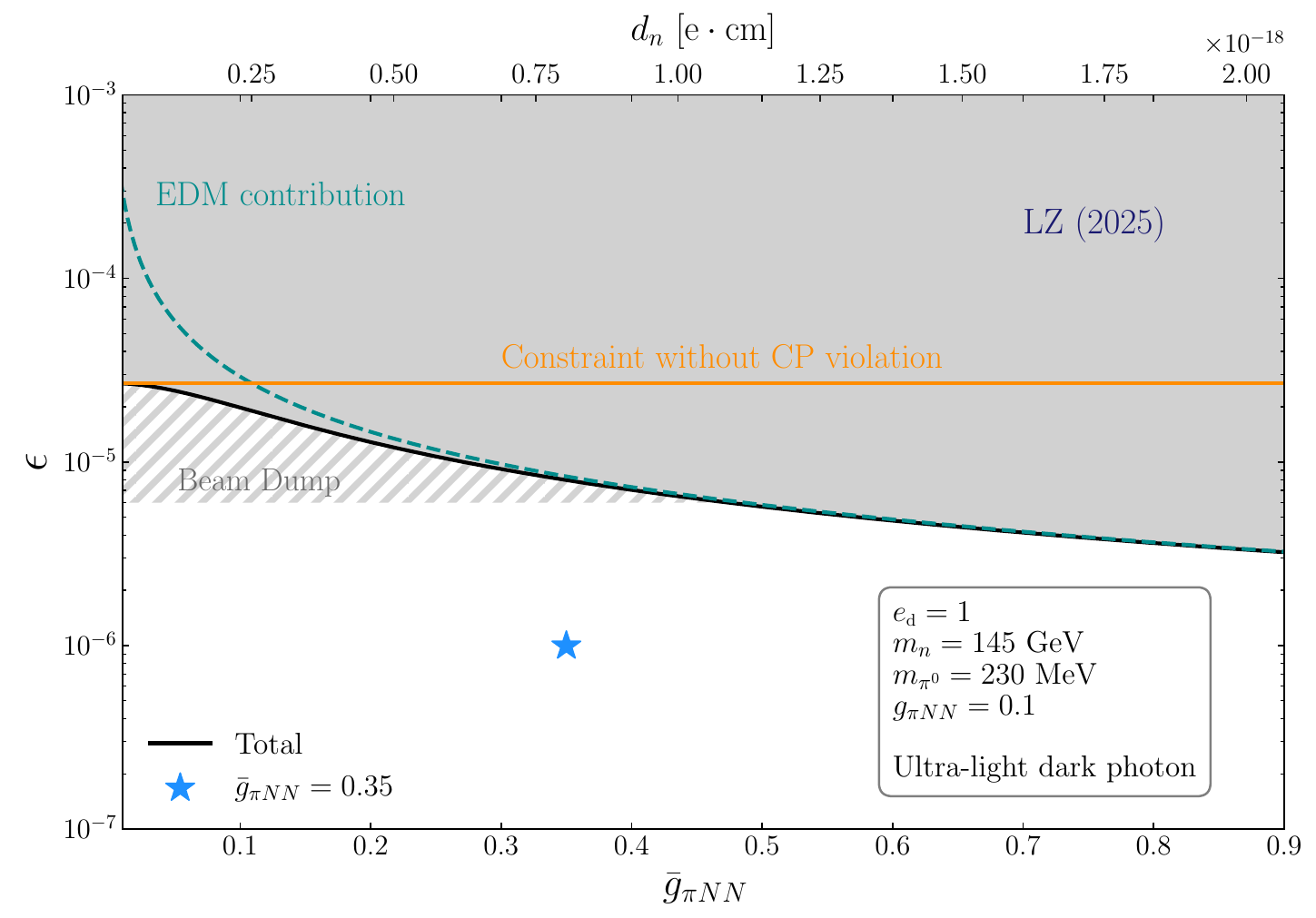}
    \caption{\textit{Top panel:} Differential event rate as a function of the nuclear recoil energy $E_R$ for the ultra-light dark photon case.
    \textit{Bottom panel:} Upper limit on the kinetic mixing $\epsilon$ in the $(\epsilon,\bar{g}_{\pi NN})$ plane, as described in the main text.}
    \label{fig:ultra_light_rates}
\end{figure}

Finally, we mention a more radical possibility, compatible with the results of Sec.~\ref{sec:sidm} and cosmological constraints: the dark sector could be completely secluded, with no interactions
with the SM (apart from  gravity). Strong interactions within the dark sector lead to thermalization of the dark quark-gluon plasma, so that the dark sector evolves with its own temperature $T_{\rm d}$. As long as the dark sector thermal bath is much colder than the SM one, $T_{\rm d}\ll T$, the abundance of dark pions is suppressed by a $(T_{\rm d}/T)^3$ factor and does not over close the universe. While such a scenario would not provide phenomenological signatures, apart from those related to DM self-interactions, it can accommodate the results of Sec.~\ref{sec:sidm}, while being consistent with cosmological constraints. Notice that, in the absence of the dark photon portal, the full baryon field $N$ in  Eq.~\eqref{eq:doublettriplet} is the DM (and not only the $n$ component), while the full pion triplet mediates DM self-scatterings, so that the computation of Sec.~\ref{sec:sidm} should be adapted to this scenario. We do not expect qualitative differences. SIDM models within a secluded dark sector have been discussed in detail in Ref.~\cite{Hambye:2019tjt}.

\section{Direct detection in the case of an ultra-light dark photon} \label{app:dd_complementary}

Following the same procedure described in the main text, we compute the corresponding differential recoil rate and derive upper limits on the kinetic  mixing $\epsilon$ in the plane $(\epsilon,\bar{g}_{\pi NN})$ for an ultra-light dark photon in Fig.~\ref{fig:ultra_light_rates}. For $m_V \lesssim 1 $ MeV, the rate is completely dominated by the $1/q^2$ enhancement. In this scenario, the relevant bounds are those coming from deviations from Coulomb's law, stellar energy-loss limits, and CMB spectral distortions~\cite{Caputo:2026pdw}. We therefore select an $\epsilon$ compatible with those constraints, together with values of $m_n$ and $m_{\pi^0}$ such that the resulting direct-detection signal is of the same order as the one that we obtain for the benchmark point used in the main text. Notice that, for such a choice of parameters, the viscosity cross section takes values in the correct ballpark to realize the SIDM scenario, see Fig.~\ref{fig:sidm_plot}.

\newpage
\bibliographystyle{JHEP}
\bibliography{ref}

\end{document}